\documentclass[onecolumn,amsmath,nofootinbib,showpacs,12pt]{revtex4-1}
\usepackage{graphicx}
\usepackage{dcolumn}
\usepackage{bm}
\usepackage{color}
\begin{document}
	%\preprint{APS/123-QED}
	%%%%%%%%%%%%%%%%%%%%%%%%
	\newcommand{\hs}{\hspace*{0.5cm}}
	\newcommand{\vs}{\vspace*{0.5cm}}
	\newcommand{\be}{\begin{equation}}
	\newcommand{\ee}{\end{equation}}
	\newcommand{\bea}{\begin{eqnarray}}
	\newcommand{\eea}{\end{eqnarray}}
	\newcommand{\ben}{\begin{enumerate}}
		\newcommand{\een}{\end{enumerate}}
	\newcommand{\bde}{\begin{widetext}}
		\newcommand{\ede}{\end{widetext}}
	\newcommand{\nn}{\nonumber}
	\newcommand{\crn}{\nonumber \\}
	\newcommand{\Tr}{\mathrm{Tr}}
	\newcommand{\non}{\nonumber}
	\newcommand{\noi}{\noindent}
	\newcommand{\al}{\alpha}
	\newcommand{\la}{\lambda}
	\newcommand{\bet}{\beta}
	\newcommand{\ga}{\gamma} 
	\newcommand{\va}{\varphi}
	\newcommand{\om}{\omega}
	\newcommand{\pa}{\partial}
	\newcommand{\+}{\dagger}
	\newcommand{\fr}{\frac}
	\newcommand{\bc}{\begin{center}}
		\newcommand{\ec}{\end{center}}
	\newcommand{\Ga}{\Gamma}
	\newcommand{\de}{\delta}
	\newcommand{\De}{\Delta}
	\newcommand{\ep}{\epsilon}
	\newcommand{\varep}{\varepsilon}
	\newcommand{\ka}{\kappa}
	\newcommand{\La}{\Lambda}
	\newcommand{\si}{\sigma}
	\newcommand{\Si}{\Sigma}
	\newcommand{\ta}{\tau}
	\newcommand{\up}{\upsilon}
	\newcommand{\Up}{\Upsilon}
	\newcommand{\ze}{\zeta}
	\newcommand{\ps}{\psi}
	\newcommand{\Ps}{\Psi}
	\newcommand{\ph}{\phi}
	\newcommand{\vph}{\varphi}
	\newcommand{\Ph}{\Phi}
	\newcommand{\Om}{\Omega}
	%%%%%%%%%%%%%%%%%%%%%%%%
	
	\title{Dark matter and flavor changing in the flipped 3-3-1 model}  
	
	\author{D. T. Huong$^1$, D. N. Dinh$^1$, L. D. Thien$^{2,3}$, and Phung Van Dong$^{3,*}$}
	\affiliation{$^1$Institute of Physics, Vietnam Academy of Science and Technology, 10 Dao Tan, Ba Dinh, Hanoi, Vietnam\\
$^2$Graduate University of Science and Technology, Vietnam Academy of Science and Technology,
18 Hoang Quoc Viet, Cau Giay, Hanoi, Vietnam
\\
$^3$Phenikaa Institute for Advanced Study and Faculty of Basic Science, Phenikaa University, Hanoi 100000, Vietnam
\\ $^*$Corresponding author. dong.phungvan@phenikaa-uni.edu.vn}
	
	\date{\today}
	
\begin{abstract}
The flipped 3-3-1 model discriminates lepton families instead of the quark ones in normal sense, where the left-handed leptons are in two triplets plus one sextet while the left-handed quarks are in antitriplets, under $SU(3)_L$. We investigate a minimal setup of this model and determine novel consequences of dark matter stability, neutrino mass generation, and lepton flavor violation. Indeed, the model conserves a noncommutative $B-L$ symmetry, which prevents the unwanted vacua and interactions and provides the matter parity and dark matter candidates that along with normal matter form gauge multiplets. The neutrinos obtain suitable masses via a type I and II seesaw mechanism. The nonuniversal couplings of $Z'$ with leptons govern lepton flavor violating processes such as $\mu \rightarrow 3e$, $\mu\rightarrow e \bar{\nu}_\mu\nu_e$, $\mu$-$e$ conversion in nuclei, semileptonic $\tau\rightarrow \mu(e)$ decays, as well as the nonstandard interactions of neutrinos with matter. This $Z'$ may also set the dark matter observables and give rise to the LHC dilepton and dijet signals.                  
	\end{abstract}
	
	\pacs{12.60.-i}
	
	\maketitle
	
	\section{Introduction}
	
The $SU(3)_C\otimes SU(3)_L\otimes U(1)_X$ (3-3-1) gauge model has been extensively studied over the last decades  \cite{Pisano:1991ee,Frampton:1992wt,Foot:1992rh,Singer:1980sw,Montero:1992jk,Foot:1994ym}. Indeed, it can resolve the profound questions of fermion generation number due to $SU(3)_L$ anomaly cancellation (see, for instance, \cite{Singer:1980sw,Frampton:1992wt}), electric charge quantization due to $SU(3)_L$ particle structure (see, for instance, \cite{Pisano:1996ht,Doff:1998we,deSousaPires:1998jc,deSousaPires:1999ca,VanDong:2005ux}), and strong CP conservation due to automatic Peccei-Quinn like symmetry (see, for instance, \cite{Pal:1994ba,Dias:2003iq,Dias:2003zt,Dong:2012bf}). Additionally, the model can supply consistent neutrino masses \cite{Tully:2000kk,Dias:2005yh,Chang:2006aa,Dong:2006mt,Dong:2008sw,Dong:2010gk,Dong:2010zu,Dong:2011vb,Boucenna:2014ela,Boucenna:2014dia,Boucenna:2015zwa,Okada:2015bxa,Pires:2014xsa,Dias:2010vt,Huong:2016kpa,Reig:2016tuk} and viable dark matter candidates~\cite{Fregolente:2002nx,Hoang:2003vj,Filippi:2005mt,deS.Pires:2007gi,Mizukoshi:2010ky,Alvares:2012qv,Profumo:2013sca,Kelso:2013nwa,daSilva:2014qba,Dong:2013ioa,Dong:2014esa,Dong:2015rka,Dong:2013wca,Dong:2014wsa,Dong:2015yra,Huong:2016ybt,Alves:2016fqe,Ferreira:2015wja,Dong:2017zxo}. The (Higgs) inflation scenarios and leptogenesis mechanism can be newly recognized in this setup, which produce the accelerated expansion of the early universe and the baryon asymmetry of the universe, respectively \cite{Huong:2015dwa,Dong:2018aak,Dong:2017ayu}.  
	
The 3-3-1 model has been well established by assigning one of quark generations to transform under $SU(3)_L$ differently from other quark generations, whereas all lepton generations transform identically under this group. This is required in order to cancel the $[SU(3)_L]^3$ anomaly \cite{Gross:1972pv,Georgi:1972bb,Banks:1976yg,Okubo:1977sc}. Recently, Fonseca and Hirsch \cite{Fonseca:2016tbn} have made an intriguing observation of a flipped 3-3-1 model, in which one of lepton generations is arranged differently from the remaining lepton generations, while all quark generations are identical under $SU(3)_L$, by contrast. This flip of quark and lepton representations converts the flavor matters in quark sector \cite{Ng:1992st,GomezDumm:1994tz,Long:1999ij,Buras:2012dp,Buras:2013dea,Gauld:2013qja,Buras:2015kwd,Dong:2015dxw} to the lepton sector. In this case, the neutral current of $Z'$ conserves quark flavors, while it violates lepton flavors. Therefore, the tree-level lepton flavor violating processes, for instance $\mu\rightarrow 3e $ and $\mu\rightarrow e \bar{\nu}_\mu\nu_e$ \cite{Fonseca:2016tbn}, exist due to the exchange of $Z'$. This flavor changing also leads to the anomalies in interaction of neutrinos with matter.   
	
As supposed in \cite{Fonseca:2016tbn}, the flipped 3-3-1 model realizes a special content of fermions. In this case, one can verify that the gravitational anomaly $[\mathrm{Gravity}]^2U(1)_X$ vanishes. This makes the model valid up to the Planck scale, where the effect of quantum gravity becomes important \cite{Delbourgo:1972xb,AlvarezGaume:1983ig}. This condition is necessary in order to ensure a manifest $B-L$ symmetry at high energy, analogous to the standard model. Additionally, $B-L$ neither commutes nor closes algebraically with $SU(3)_L$, similar to the electric charge. The algebraic closure condition results in a complete gauge symmetry $SU(3)_C\otimes SU(3)_L\otimes U(1)_X\otimes U(1)_N$, where $N$ determines $B-L$, in the same situation that $X$ defines the electric charge \cite{Dong:2013wca,Dong:2014wsa,Huong:2015dwa,Dong:2015yra,Huong:2016ybt,Alves:2016fqe,Dong:2018aak}. The $B-L$ breaking leads to a matter parity which characterizes and stabilizes dark matter candidates. It is noteworthy that dark matter is unified with normal matter in gauge multiplets due to the noncommutativity of $B-L$ symmetry. Additionally, the matter parity cures the unwanted vacua and interactions, which otherwise imply large flavor-changing neutral currents (FCNCs) in the lepton sector. The type~I seesaw can be realized due to the $B-L$ dynamics. However, this model contains naturally type II and III seesaws too.     
	
The rest of this work is organized as follows. In Sec. \ref{model} we examine the flipped 3-3-1 model when imposing the $B-L$ symmetry and matter parity. The dark matter candidates are identified. The masses of lepton sextet including neutrinos are obtained. The gauge sector is diagonalized. In Sec.~\ref{fcnc} we determine the tree-level FCNC coupled to $Z'$. The setup also implies the tree-level FCNC coupled to the standard model Higgs boson, but it is subleading and neglected. In Sec. \ref{pheno}, the lepton flavor violating processes, the nonstandard interactions of neutrinos with matter, the LHC dilepton and dijet searches, and the dark matter observables are obtained. Finally, we conclude this work in Sec. \ref{conl}.                         
	
\section{\label{model} Novel features of the model}

\subsection{Proposal}

As stated, the 3-3-1 gauge symmetry is given by \be SU(3)_C\otimes
SU(3)_L\otimes U(1)_X,\ee where the first factor is the usual QCD group, while the last two are a nontrivial extension of the electroweak group. The electric charge and hypercharge are embedded as
\be Q=T_3+\beta T_8 +X,\hs Y=\beta T_8 + X,\ee where $T_n$ ($n=1,2,3,...,8$) and $X$ are $SU(3)_L$ and $U(1)_X$ generators, respectively. The coefficient $\beta$ determines the electric charge of new particles, which is arbitrary on the theoretical ground and independent of all the anomalies.  
	
The new observation \cite{Fonseca:2016tbn} is that the $[SU(3)_L]^3$ anomaly ($\mathcal{A}$) induced by a fermion sextet is related to that by a fermion triplet as $\mathcal{A}(6)=7\mathcal{A}(3)$, where the color number is not counted. If one puts a lepton generation in a sextet and two other lepton generations in triplets, the anomaly contributed by the three lepton generations equals $9\mathcal{A}(3)$. This cancels the contribution of three quark generations arranged in antitriplets, since $\mathcal{A}(3^*)=-\mathcal{A}(3)$ and quarks have three colors. In general, it is proved that the generation number must be a multiple of three in order to embed left-handed fermion doublets in $SU(3)_L$ representations while keeping right-handed fermions as singlets similar to the standard model, which provides a partial solution to the flavor question. 

That said, the left-handed fermion representations under $SU(3)_L$ are generally given by 
	\bea \psi_{1L} &=& \left(  \begin{array}{ccc} % or pmatrix or bmatrix or Bmatrix
		%or ...
		\xi^{-q}_1 & \fr{1}{\sqrt{2}}\xi^{-1-q}_2 & \fr{1}{\sqrt{2}} \nu_1 \\
		\fr{1}{\sqrt{2}} \xi^{-1-q}_2 & \xi^{-2-q}_3 & \fr{1}{\sqrt{2}} e_1 \\
		\fr{1}{\sqrt{2}} \nu_1 & \fr{1}{\sqrt{2}} e_1 & k_1^q \\
	\end{array}\right)_L\sim 3,\\
	\psi_{\al L} &=& \left(
	\begin{array}{c}
		\nu_{\al }\\
		e_{\al }\\
		k^q_{\al}\\
	\end{array}
	\right)_L\sim 3,\hs
	Q_{aL} = \left(
	\begin{array}{c}
		d_a\\
		-u_a\\
		j^{-q-1/3}_a\\
	\end{array}
	\right)_L\sim 3^*,\eea plus right-handed fermion singlets, $\xi_{aR},\ e_{aR},\ k_{aR},\ u_{aR},\ d_{aR }$, and $j_{aR}$. The generation indices are $\al =2,3$ and $a=1,2,3$. The new fields $\xi_{1,2,3}$, $k_a$, and $j_a$ possess electric charges as superscripted, which depend on a basic electric charge parameter $q=-(1+\sqrt{3}\beta)/2$. It is easily checked that all the other anomalies vanish. An alternative version can be proposed, such that a lepton generation is in antisextet, two other lepton generations in antitriplets, and all the quark generations in triplets, under $SU(3)_L$. However, this modification should be equivalently physical to the original proposal, which will be skipped. 
	
The previous study \cite{Fonseca:2016tbn} considered a special version, given that $q=-1$ or $\beta=1/\sqrt{3}$~\footnote{The 3-3-1 model with $\beta=1/\sqrt{3}$ has a Landau pole larger than the Planck scale, as expected \cite{Dias:2004dc,Dias:2004wk}.}. In this case, $\xi^0_{2R}$ is a 3-3-1 singlet which can be omitted as $\nu_{aR}$ are. Additionally, $\xi^{-}_{3R}$ can be discarded while $\xi^{-}_{3L}$ is replaced by $(\xi^+_{1R})^c=\xi^-_{1L}$. This restriction results in the most economical model, explicitly written under the 3-3-1 symmetry as    
	\bea \psi_{1L}&=& \left(  \begin{array}{ccc} % or pmatrix or bmatrix or Bmatrix
		%or ...
		\xi^{+} & \fr{1}{\sqrt{2}}\xi^{0} & \fr{1}{\sqrt{2}} \nu_1 \\
		\fr{1}{\sqrt{2}} \xi^{0} & \xi^{-} & \fr{1}{\sqrt{2}} e_1 \\
		\fr{1}{\sqrt{2}} \nu_1 & \fr{1}{\sqrt{2}} e_1 & E_1 \\
	\end{array}\right)_L\sim \left(1,6,-\fr{1}{3}\right),\\
	\psi_{\al L}&=&\left(
	\begin{array}{c}
		\nu_{\al }\\
		e_{\al }\\
		E_{\al}\\
	\end{array}
	\right)_L\sim \left(1,3,-\fr{2}{3}\right),\\
	e_{aR} &\sim& (1,1,-1),\hs E_{aR}\sim (1,1,-1),\\
	Q_{aL}&=&\left(
	\begin{array}{c}
		d_a\\
		-u_a\\
		U_a\\
	\end{array}
	\right)_L\sim \left(3,3^*,\fr{1}{3}\right),\\ 
	u_{aR} &\sim& (3,1,2/3),\hs d_{aR }\sim (3,1,-1/3),\hs U_{aR}\sim
	(3,1,2/3),\eea where we relabel $k$ as $E$ and $j$ as $U$, since they have the same electric charge as of $e$ and $u$, respectively. The subscripts of $\xi$ are omitted without confusion. The authors of \cite{Fonseca:2016tbn} pointed out that since the model cooperates a real triplet $\xi$ under $SU(2)_L$, it can be made heavy in order to keep the model phenomenologically viable. Further, we will show that the simplest version provides dark matter candidates as well as neutrino masses naturally. Otherwise, for $q\neq -1$, the fermions $\xi$ must have right-handed components and gain light Dirac masses in the weak scale, which are strongly disfavored by the electroweak data.     
	
The scalar content responsible for symmetry breaking and mass generation is given by 
	\bea &&\eta=\left(
	\begin{array}{c}
		\eta^0_1\\
		\eta^-_2\\
		\eta^-_3\\
	\end{array}
	\right)\sim (1,3,-2/3),\hs \rho=\left(
	\begin{array}{c}
		\rho^+_1\\
		\rho^0_2\\
		\rho^0_3\\
	\end{array}
	\right)\sim (1,3,1/3),\\
	&& \chi=\left(
	\begin{array}{c}
		\chi^+_1\\
		\chi^0_2\\
		\chi^0_3\\
	\end{array}
	\right)\sim (1,3,1/3),\hs S=\left(\begin{array}{ccc}
		S^{++}_{11} & \fr{1}{\sqrt{2}} S^{+}_{12} & \fr{1}{\sqrt{2}} S^+_{13}\\
		\fr{1}{\sqrt{2}}S^+_{12} & S^0_{22} & \fr{1}{\sqrt{2}} S^0_{23}\\
		\fr{1}{\sqrt{2}}S^+_{13} & \fr{1}{\sqrt{2}} S^0_{23} & S^0_{33}\\
	\end{array}\right)\sim (1,6,2/3).\eea Note that $\rho$ and $\chi$ are identical under the gauge symmetry, but distinct under the $B-L$ charge, as shown below.

\subsection{Dark matter}	
	
First note that the electric charge neither commutes nor closes algebraically with $SU(3)_L$, since $[Q,T_n]\neq 0$ for $n=1,2,4,5$ and $\mathrm{Tr} Q\neq 0$ for various particle multiplets. This property applies for every 3-3-1 model, including the standard model with $SU(2)_L$.  

Similarly to the electric charge, $B-L$ neither commutes nor closes algebraically with $SU(3)_L$, which differs from the case of the standard model. Indeed, the standard model conserves $U(1)_{B-L}$, which follows that $[B-L](\xi^+)=[B-L](\xi^-)=[B-L](\xi^0)\equiv n$ for a $SU(2)_L$ triplet. Using the condition $[B-L](\xi^+)=-[B-L](\xi^-)$, we obtain $n=0$. Supposing that the 3-3-1 model conserves $B-L$ charge, we find $B-L=\mathrm{diag}(0,0,0,-1,-1,-2)$ for the sextet $(\xi^+\ \xi^0\ \xi^-\ \nu_1\ e_1\ E_1)_L$ and $B-L=\mathrm{diag}(-1,-1,-2)$ for the triplets $(\nu_{\al}\ e_{\al}\ E_{\al})_L$. This implies that $[B-L,T_n]\neq 0$ for $n=4,5,6,7$ and $\mathrm{Tr}[B-L]\neq 0$, as expected. 

The requirement of algebraic closure between $B-L$ and $SU(3)_L$ results in an extra $U(1)_N$ symmetry, such that     
\be B-L=\fr{2}{\sqrt{3}} T_8+N,\ee 
where $N$ determines $B-L$ in the same situation that $X$ does so for $Q$ \cite{Dong:2013wca,Dong:2014wsa,Huong:2015dwa,Dong:2015yra,Huong:2016ybt,Alves:2016fqe,Dong:2018aak}. This way leads to the group structure $SU(3)_C\otimes SU(3)_L\otimes U(1)_X\otimes U(1)_N$, called 3-3-1-1. Additionally, $N$ and $B-L$ are gauged charges since they are related to the gauged charge $T_8$. The 3-3-1-1 gauge theory requires $\nu_{aR}$ in addition to the existing fermions in order to cancel the $B-L$ anomalies as well as a scalar field $\phi$ that couples to $\nu_R\nu_R$ and necessarily breaks $U(1)_N$. The $N$-charges of multiplets are summarized in Tab. \ref{tbadd0}
\begin{table}
\begin{tabular}{lcccccccccccccc}
\hline\hline
Multiplet & $\psi_{1L}$ & $\psi_{\al L}$ & $Q_{aL}$ & $e_{aR}$ & $E_{aR}$ & $u_{aR}$ & $d_{aR}$ & $U_{aR}$ & $\eta$ & $\rho$ & $\chi$ & $S$ & $\nu_{aR}$ & $\phi$\\
\hline
$N$ & $-2/3$ & $-4/3$ & $2/3$ & $-1$ & $-2$ & $1/3$ & $1/3$ & $4/3$ & $-1/3$ & $-1/3$ & $2/3$ & $4/3$ & $-1$ & $2$\\
\hline\hline
\end{tabular}
\caption{\label{tbadd0} $N$-charge of the model's multiplets.}
\end{table} 

Assuming that $U(1)_N$ is broken at high energy due to $\phi$, the heavy particles such as the $U(1)_N$ gauge $(C)$ and Higgs ($\phi$) bosons as well as right-handed neutrinos ($\nu_{aR}$) are all integrated out. The imprint at low energy is only the 3-3-1 model, conserving the matter parity as residual gauge symmetry,\be 
	W_P=(-1)^{3(B-L)+2s}=(-1)^{2\sqrt{3}(T_8+N)+2s},\ee which is defined by the vacuum of the mentioned $U(1)_N$ breaking field \cite{Dong:2015yra}. 
	
Above, ``$W$'' means the fields that have ``wrong'' $B-L$ charge in comparison to the standard model which transform nontrivially under the matter parity. They are collected in Tab. \ref{tbadd1}. The remaining fields have $W_P=1$, called normal fields. 
\begin{table}
\begin{tabular}{lccccccccccccc}
\hline\hline
Field & $\xi^{+}$ & $\xi^{0}$ & $\xi^{-}$ & $E^-$ & $U^{2/3}$ & $X^+$ & $Y^0$ & $\eta^-_3$ & $\rho^0_3$ & $\chi^+_{1}$ & $\chi^0_2$ & $S^+_{13}$ & $S^0_{23}$\\
\hline  
$B-L$ & 0 & 0 & 0 & $-2$ & $4/3$ & $1$ & $1$ & $-1$ & $-1$ & $1$ & 1 & 1 & 1
\\
$W_P$ & $-1$ & $-1$ & $-1$ & $-1$ & $-1$ & $-1$ & $-1$ & $-1$ & $-1$ & $-1$ & $-1$ & $-1$ & $-1$\\
\hline\hline
\end{tabular}
\caption{\label{tbadd1} Nontrivial matter parity and $B-L$ charge.}
\end{table} 

Because the matter parity is conserved, the lightest $W$-particle (LWP) is stabilized, responsible for dark matter. The dark matter candidates include a fermion $\xi^0$, a vector $Y^0$, and a combination of $\rho^0_3$, $\chi^0_2$, and $S^0_{23}$. Due to the gauge interaction, $Y^0$ annihilates completely into the standard model particles. The realistic candidates that have correct abundance are only the fermion or scalar, as shown below.   
    
\subsection{Lagrangian}

Hereafter, we consider the theory at low energy that includes only the light fields. The contribution of heavy fields ($C,\phi,\nu_{aR}$) is separately mentioned if necessary. 

The total Lagrangian consists of \be
	\mathcal{L}=\mathcal{L}_{\mathrm{kinetic}}+\mathcal{L}_{\mathrm{Yukawa}}-V,\ee
where the first part composes the kinetic terms plus gauge interactions. 
	
The second part includes Yukawa interactions, obtained by 
	\bea \mathcal{L}_{\mathrm{Yukawa}}&=& h^{e}_{\al a} \bar{\psi}_{\al L} \rho
	e_{a R} +h^{E}_{\al a}\bar{\psi}_{\al L}\chi
	E_{aR}+h_{1a}^{E}\bar{\psi}_{1L}S E_{aR}+h^\xi
	\bar{\psi}^c_{1L}\psi_{1L}S\crn
	&& +h_{ab}^{u}\bar{Q}_{a L}\rho^* u_{bR}
	+h_{ab}^d \bar{Q}_{aL}\eta^* d_{bR}
	+h_{ab}^{U}\bar{Q}_{a L}\chi^* U_{bR}+H.c.\eea
Note that the unwanted Yukawa interactions are 
\bea \mathcal{L}\!\!\!\!/_{\mathrm{Yukawa}}&=&s^{e}_{\al a}\bar{\psi}_{\al L}\chi
	e_{aR}+s_{1a}^{e}\bar{\psi}_{1L}S e_{aR} +s^{E}_{\al a}
	\bar{\psi}_{\al L} \rho E_{a R}\crn
	&&+s_{ab}^{u}\bar{Q}_{a L}\chi^* u_{bR} +s_{ab}^{U}\bar{Q}_{a L}\rho^* U_{bR}+H.c.,\eea which are suppressed due to the $U(1)_N$ symmetry or matter parity at low energy. In the ordinary 3-3-1 models, they are present, characterizing an approximate $B-L$ symmetry. In other words, they should be small in comparison to the normal couplings $s\ll h$, respectively.  
  
The last part is the scalar potential, 
	\bea V&=&\mu_\eta^2 \eta^\dag \eta+\mu_{\rho}^2
	\rho^\dag \rho+\mu_{\chi}^2 \chi^\dag \chi+\mu_S^2 \Tr(S^\dag S)\crn
	&&+\la_{\eta}(\eta^\dag
	\eta)^2+\la_{\rho}(\rho^\dag \rho)^2+\la_\chi \left(\chi^\dag \chi\right)^2+\la_{1S}\Tr^2(S^\dag
	S)+\la_{2S}\Tr(S^\dag S)^2
	\nonumber \\  &&+\la_{\eta \rho}(\eta^\dag
	\eta)(\rho^\dag \rho)+\la_{\chi \eta}(\chi^\dag \chi)(\eta^\dag \eta)+\la_{\chi \rho}(\chi^\dag
	\chi)(\rho^\dag \rho)\crn
	&&+\la_{\eta
		S}(\eta^\dag \eta) \Tr(S^\dag S) +
	\la_{ \rho S}(\rho^\dag \rho) \Tr(S^\dag S)+\la_{\chi S}(\chi^\dag \chi) \Tr(S^\dag S)\crn
	&&+\la_{\eta \rho}^\prime(\eta^\dag \rho)(\rho^\dag
	\eta)+\la^\prime_{\chi
		\eta}(\chi^\dag\eta)(\eta^\dag \chi)+\la^\prime_{\chi \rho} (\chi^\dag
	\rho)(\rho^\dag \chi)\nonumber \\ &&+\la_{\chi S}^\prime (\chi^\dag S)(S^\dag \chi)+\la_{\eta
		S}^\prime (\eta^\dag S)(S^\dag \eta)+\la_{\rho S}^\prime (\rho^\dag S)(S^\dag
	\rho)\crn
	&&+
	\left(\mu \eta \rho \chi+\mu^\prime \chi^T S^* \chi +H.c. \right)\eea Here the couplings $\la$'s are dimensionless, while the parameters $\mu$'s have mass dimension. The unwanted interactions include
	\be V\!\!\!\!/= f^2 \rho^\dagger \chi + f' \rho^T S^* \chi + f''\rho^T S^* \rho +H.c.,\ee which are suppressed by the $U(1)_N$ symmetry or matter parity at low energy, where $f$'s are mass parameters analogous to $\mu$'s. 

\subsection{Neutrino mass}

First note that due to the matter parity conservation, only the even scalars develop vacuum expectation values (VEVs), such that 
\bea &&\langle \eta\rangle =\fr{1}{\sqrt{2}}\left(
	\begin{array}{c}
		u\\
		0\\
		0\\
	\end{array}
	\right),\hs \langle\rho\rangle =\fr{1}{\sqrt{2}}\left(
	\begin{array}{c}
		0\\
		v\\
		0\\
	\end{array}
	\right),\\
	&& \langle \chi\rangle =\fr{1}{\sqrt{2}}\left(
	\begin{array}{c}
		0\\
		0\\
		w\\
	\end{array}
	\right),\hs \langle S\rangle =\fr{1}{\sqrt{2}}\left(\begin{array}{ccc}
		0 & 0 & 0\\
		0 & \kappa & 0\\
		0 & 0 & \La \\
	\end{array}\right).\eea
	
Substituting the VEVs into the Yukawa Lagrangian, the quarks and exotic leptons gain suitable masses as follows
\bea &&[m_u]_{ab}=\fr{h^u_{ab}}{\sqrt{2}}v,\hs [m_d]_{ab}=-\fr{h^d_{ab}}{\sqrt{2}}u,\hs [m_U]_{ab}=-\fr{h^U_{ab}}{\sqrt{2}}w,\\
&& m_\xi = -\sqrt{2}h^\xi \La,\hs [m_E]_{1b}=-\fr{h^E_{1b}}{\sqrt{2}}\La,\hs [m_E]_{\al b}=-\fr{h^E_{\al b}}{\sqrt{2}}w. \eea 
	
The ordinary leptons obtain masses
\be [m_e]_{\al b}=-\fr{h^e_{\al b}}{\sqrt{2}}v,\hs [m_{\nu}]_{11}=\sqrt{2}\kappa h^\xi.\ee Note that the constraint from the $\rho$ parameter implies $\kappa\lesssim\mathcal{O}(1)$ GeV. Hence, a small mixing between $\xi^\pm$ and $E^\pm_a$ proportional to $\kappa$ can be neglected, since $\kappa\ll w,\La$, where the last two VEVs are in the TeV scale. The fields $\mu$ and $\tau$ get desirable masses. However, the electron and  last two neutrinos have vanishing masses, which are inconsistent.      

The electron mass vanishes similarly to the original study, which can be radiatively induced \cite{Fonseca:2016tbn}. An extra remark is that the interaction $\psi_{1L}\psi_{1L}S\supset (\xi^+\ \xi^0\ \xi^-)(\nu_{1L}\ e_{1L})(S^+_{13}\ S^0_{23})$ provides the need for a type III seesaw, where the mediator is a heavy fermion triplet, $\xi$. But the relevant neutrino ($\nu_1$) mass generated is zero due to $\langle S^0_{23}\rangle=0$ resulting from the matter parity conservation, which does not change the above value of a type II seesaw.   

As mentioned, the heavy fields $\phi,\nu_R$ are present and can imply neutrino masses via 
\be \mathcal{L}_{\nu}= h^\nu_{\al b} \bar{\psi}_{\al L}\eta \nu_{b R}+\fr 1 2 h^R_{ab}\bar{\nu}^c_{aR}\nu_{bR}\phi+H.c.\ee We achieve Dirac masses $[m^D_\nu]_{\al b }=-h^\nu_{\al b}u/\sqrt{2}$ and Majorana masses $[m^R_\nu]_{ab}=-h^R_{ab}\langle \phi\rangle$. Because of $u\ll \langle \phi\rangle $, the observed neutrinos $\sim \nu_L$ gain masses via a type I seesaw, by 
\be [m_\nu]_{\al\beta}\simeq -[m^D_\nu (m^R_\nu)^{-1}(m^D_\nu)^T]_{\al\beta}=h^\nu_{\al a} (h^R)^{-1}_{ab}(h^\nu)^T_{b\beta}\fr{u^2}{2\langle \phi\rangle}\sim \fr{u^2}{\langle \phi\rangle}.\ee Fitting the data $m_\nu\sim 0.1$ eV, we obtain $\langle \phi\rangle\sim [(h^\nu)^2/h^R] 10^{14}\ \mathrm{GeV}$, since $u$ is proportional to the weak scale. Given that $h^\nu,h^R\sim 1$, one has $\langle \phi \rangle \sim 10^{14}$ GeV, close to the grand unification scale. The right-handed neutrinos $\nu_{aR}$ have masses in this scale.

It is clear that two neutrinos $\nu_{2,3L}$ achieve masses via the type I seesaw with the corresponding mixing angle $\theta_{23}$ comparable to the data, while the neutrino $\nu_{1L}$ has a mass (which one sets $h^\xi\kappa\sim 0.1$ eV) via the type II seesaw and does not mix with $\nu_{2,3L}$. The mixing angles $\theta_{12}$ and $\theta_{13}$ can be induced by an effective interaction, such as 
\be \mathcal{L}_{\mathrm{mix}}=\fr{h^\nu_{1 \beta}}{\mathcal{M}^2}\bar{\psi}^c_{1L}\psi_{\beta L}\rho \eta^*\phi+H.c.,\ee where $\mathcal{M}$ is the new physics scale which can be fixed at $\mathcal{M}=\langle \phi\rangle$. The mass matrix of observed neutrinos is corrected by
\be [m_\nu]_{1\beta}=-h^\nu_{1\beta}\fr{uv}{\langle \phi\rangle}\sim \fr{uv}{\langle \phi\rangle},\ee where $v$ is proportional to the weak scale like $u$. These elements can generate appropriate mixing angles for $\theta_{12,13}$. In this case, the neutrino mass matrix is generic and small.         	
		
\subsection{Gauge sector} 
	
The mass Lagrangian of gauge bosons is given by  
	\bea 
	\mathcal{L}\supset \sum_{\Phi=\eta,\rho,\chi,S}\left(D_\mu\langle\Phi\rangle \right)^\dagger
	\left(D^{\mu}\langle \Phi\rangle\right), 
	\eea
	where $D_\mu=\pa_\mu + i g_s t_n G_{n\mu}+i g T_n A_{n \mu}+ig_X X B_\mu$ is covariant derivative, in which $(g_s,g,g_X)$, $(t_n,T_n,X)$, and $(G_n,A_n,B)$ correspond to the coupling constants, generators, and gauge bosons of the 3-3-1 groups, respectively\footnote{Note that the $U(1)_N$ part and its scalar were integrated out.}. It acts as $D_\mu\langle \Phi\rangle =ig(T_n A_{n\mu}+t_X X_\Phi B_\mu)\langle \Phi\rangle$ for a triplet and $D_\mu\langle S\rangle = ig(T_n A_{n\mu} \langle S\rangle + \langle S\rangle T_n^T A_{n\mu} +t_X X_S B_\mu \langle S\rangle)$ for a sextet, where $T_n= \fr 1 2 \la_n$ are Gell-Mann matrices and $t_X=g_X/g$.  
	
Define the non-Hermitian gauge bosons,
	\bea
	W^\pm =\fr{1}{\sqrt{2}}\left(A_{1}\mp iA_{2} \right), \hs
	X^\pm =\fr{1}{\sqrt{2}}\left(A_{4}\mp iA_{5} \right), \hs Y^{0,0*} =\fr{1}{\sqrt{2}}\left(A_{6}\mp iA_{7} \right).
	\eea
They are mass eigenstates by themselves with corresponding masses, 
\be m^2_W\simeq \fr{g^2}{4}(u^2+v^2),\hs m^2_X=\fr{g^2}{4}(u^2+w^2+2\La^2),\hs m^2_Y\simeq \fr{g^2}{4}(v^2+w^2+2\La^2).\ee Because of the matter parity conservation, there is no mixing between $W$ and $X$ as well as $A_6$ and neutral gauge bosons. Because the type II seesaw requires an infinitesimal $\kappa$, its correction to $m_W$ and $m_Y$ has been suppressed.   
		
The neutral gauge bosons $(A_3,A_8,B)$ mix via a 3$\times$3 mass matrix. This yields the massless photon field,
	\bea
	A =s_W A_3+c_W \left( \fr{t_W}{\sqrt{3}}
	A_8+\sqrt{1-\fr{t^2_W}{3}} B \right), 
	\eea
	where $s_W=e/g=\sqrt{3}t_X/\sqrt{3+4t^2_X}$ is the sine of the Weinberg's angle \cite{VanDong:2005pi}. As usual, we define the remaining neutral fields orthogonal to $A$ as 
	\bea
	Z &&=c_W A_3-s_W \left( \fr{t_W}{\sqrt{3}}
	A_8+\sqrt{1-\fr{t^2_W}{3}} B_\mu \right), \\
	Z^\prime &&=\sqrt{1-\fr{t^2_W}{3}} A_{8}-\fr{t_W}{\sqrt{3}}
	B.
	\eea
In the new basis $(A,Z,Z')$, there is only $Z$-$Z'$ mixing via a $2\times 2$ mass matrix, while $A$ is decoupled. This mass matrix yields eigenstates
\be Z_1=c_\varphi Z- s_\varphi Z',\hs Z_2 = s_\varphi Z+c_\varphi Z',\ee with masses  	
	\bea
	m^2_{Z_1} &\simeq& \fr{g^2}{4c^2_W}\left(u^2+v^2\right),\\
	m^2_{Z_2} &\simeq& \fr{g^2}{4(3-t^2_W)}\left[(1+t^2_W)^2 u^2+(1-t^2_W)^2 v^2+4(w^2+4\La^2)\right],\eea and the mixing angle
	\be t_{2\varphi}\simeq\fr{\sqrt{3-4s^2_W}}{2c^4_W} \fr{u^2-c_{2W}v^2}{w^2+4\La^2}.\ee
	
Since $\kappa$ is tiny, its contribution to the $\rho$ parameter is neglected. The deviation of the $\rho$ parameter from the standard model prediction is due to the $Z$-$Z'$ mixing, obtained by 
\be \Delta\rho \simeq \fr{(u^2-c_{2W}v^2)^2}{4c^4_W(u^2+v^2)(w^2+4\La^2)}.\ee  

From the $W$ mass, we derive $u^2+v^2=(246\ \mathrm{GeV})^2$. From the global fit, the PDG Collaboration extracts the $\rho$ deviation as $\Delta\rho=0.00039\pm 0.00019$, which is $2\sigma$ above the standard model prediction \cite{Tanabashi:2018oca}. We contour $\Delta\rho$ as a function of $u=0$--246 GeV, since $v$ is related to $u$. Generally for the whole $u$ range, the new physics scales are bounded by $\sqrt{w^2+4\La^2}\sim$ 5--7~TeV \cite{Dong:2015jxa}. However, there is a regime localized at $u=\sqrt{c_{2W}}v\simeq 145$ GeV, where both $\Delta\rho$ and $\varphi$ vanish, i.e. the new physics is always decoupled when $w,\La$ tend to zero. Therefore, we can close the 3-3-1 symmetry at the weak scale in this regime \cite{Dias:2005xj,Dias:2006ns}.       
	
\section{\label{fcnc}FCNC} 
	
Because the first lepton generation transforms differently from the last two lepton generations under $SU(3)_L$, there exist FCNCs at the tree-level associated with leptons. Note that the FCNCs conserve quark flavors, in contrast to the ordinary 3-3-1 models. 

Indeed, the neutral currents of fermions
depend on the Cartan (diagonal) charges $T_{3,8}$ and $X=Q-T_3-T_8/\sqrt{3}$ as follows
\bea \mathcal{L} \supset \bar{F}i\ga^\mu D_\mu F\supset -g\bar{F}\ga^\mu[T_3A_{3\mu}+T_8A_{8\mu}+t_X (Q-T_3-T_8/\sqrt{3}) B_\mu]F,\eea where $F$ runs over fermion multiplets. It is easily verified that all the quarks $u_{aL}$, $u_{aR}$, $d_{aL}$, $d_{aR}$, $U_{aL}$, and $U_{aR}$ as well as the right-handed leptons $e_{aR}$ and $E_{aR}$ correspondingly do not flavor-change. Also, the terms of $T_3$ and $Q$ conserve all fermion flavors. 
There remains
\bea \mathcal{L} \supset -g\bar{\psi}_{aL}\ga^\mu T_8 \psi_{aL}\left(A_{8\mu}-\fr{t_X}{\sqrt{3}}B_\mu\right)=-\fr{g}{\sqrt{1-t^2_W/3}}\bar{\psi}_{aL}\ga^\mu T_8 \psi_{aL}Z'_\mu,\eea where note that $T_8\psi_{\al L}=\fr 1 2 \la_8 \psi_{\al L}$, while $T_8\psi_{1 L}=\fr 1 2 \la_8 \psi_{1 L}+ \psi_{1L}\fr 1 2 \la_8$ and the corresponding interaction is traced. We obtain
\be \mathcal{L}\supset -\fr{g}{2\sqrt{3-t^2_W}}(\bar{\nu}_L \ga^\mu T_\nu \nu_L +\bar{e}_L \ga^\mu T_e e_L + \bar{E}_L \ga^\mu T_E E_L)Z'_\mu, \ee where $T_\nu=T_e=\mathrm{diag}(-1,1,1)$ and $T_E=\mathrm{diag}(-4,-2,-2)$. We commonly denote $l=\nu$ or $e$ or $E$ and change to the mass basis $l_{L,R}=V_{lL,R}l'_{L,R}$. The relevant Lagrangian is 
\bea \mathcal{L} \supset -\fr{g}{2\sqrt{3-t^2_W}} \bar{l}'_L\ga^\mu (V^\dagger_{lL}T_l V_{lL} ) l'_LZ'_\mu\equiv \Ga^{l'Z'}_{ij}\bar{l}'_{iL}\ga^\mu l'_{jL}Z'_\mu,\label{FCNC2}\eea      
where \bea \Ga^{l'Z'}_{ij}&=&-\fr{g}{2\sqrt{3-t^2_W}}(V^\dagger_{lL}T_l V_{lL})_{ij}=
\fr{g}{\sqrt{3-t^2_W}}(V^*_{lL})_{1i}(V_{lL})_{1j}, 
\eea which takes the same form for both $l=\nu,e$ and $l=E$ as well as flavor-changes for $i\neq j$.
	
Let us recall that all quark generations transform universally under the 3-3-1 group, which yields the flavor-conserved current with $Z'$ as 
	\bea
	\mathcal{L}&\supset& -\fr{g(2+c_{2W})}{6c_W\sqrt{1+2c_{2W}}}\left(\bar{u}_L
	\ga^\mu u_L+\bar{d}_L \ga^\mu d_L  \right)Z^\prime_\mu\crn
	&& -\frac{gs_W^2}{3c_W\sqrt{1+2c_{2W}}}\left( 2\bar{u}_R \ga^\mu u_R-\bar{d}_R
	\ga^\mu d_R\right)Z^\prime_\mu,\label{FCNC3}\eea where we denote physical eigenstates for up-quarks as $u=(u\ c\ t)^T$, down-quarks as $d=(d\ s\ b)^T$, and commonly $q=u$ or $d$ in further investigation.   

Integrating out the heavy field, $Z^\prime$, from (\ref{FCNC2}) and/or (\ref{FCNC3}), we obtain an effective Lagrangian that sums over six-dimensional interactions (operators) relevant to the standard model fermions at the tree-level, such that  
	\be
	-\fr{\Ga^{lZ^\prime}_{\al \beta}\Ga^{lZ^\prime}_{\ga
			\delta}}{m_{Z^\prime}^2}\left(\bar{l}_{\al} \ga^\mu P_L l_\beta 
	\right)\left(\bar{l}_\ga  \ga_\mu P_L l_{\delta}
	\right),\label{dh1}\ee	
	\be -\fr{\Ga^{lZ^\prime}_{\al \beta}}{m_{Z^\prime}^2}\left(\fr{g
		s_W^2}{c_W\sqrt{1+2c_{2W}}}\right)\left(\bar{l}_{\al} \ga^\mu P_L l_\beta 
	\right)\left(\bar{l}_\delta \ga_\mu P_R l_\delta \right),\label{dh2}\ee 	
	\be -\fr{\Ga^{lZ^\prime}_{\al \beta}\Ga^{\nu Z^\prime}_{\ga
			\delta}}{m_{Z^\prime}^2}\left(\bar{\nu}_\ga \ga_\mu  P_L \nu_{\delta}
	\right)\left(\bar{l}_{\al} \ga^\mu P_L l_\beta  \right),\label{dh3}\ee	
	\be -\fr{\Ga^{\nu
			Z^\prime}_{\al \beta}}{m_{Z^\prime}^2}\left(\fr{g
		s_W^2}{c_W\sqrt{1+2c_{2W}}}\right)\left(\bar{\nu}_\al  \ga_\mu P_L \nu_{\beta}
	\right)\left(\bar{l}_{\delta} \ga^\mu P_R l_\delta  \right),\label{dh4}\ee	
	\be +\fr{\Ga^{\nu Z^\prime}_{\al
			\beta}}{m_{Z^\prime}^2}\fr{g(2+c_{2W})}{6c_W\sqrt{1+2c_{2W}}}\left(\bar{\nu}_{\al}
	\ga^\mu P_L \nu_\beta  \right)
	\left(\bar{q} \ga_\mu P_L q \right),\label{dh5}\ee	
	\be +\fr{\Ga^{\nu Z^\prime}_{\al
			\beta}}{m_{Z^\prime}^2}\frac{gs_W^2}{3c_W\sqrt{1+2c_{2W}}}\left(\bar{\nu}_{\al}
	\ga^\mu P_L \nu_\beta  \right)
	\left(\eta^q \bar{q}  \ga_\mu P_R q \right),\label{dh6}
	\ee 	
	\be +\fr{\Ga^{lZ^\prime}_{\al
			\beta}}{m_{Z^\prime}^2}\fr{g(2+c_{2W})}{6c_W\sqrt{1+2c_{2W}}}\left(\bar{l}_{\al}
	\ga^\mu P_L l_\beta  \right)
	\left(\bar{q} \ga_\mu P_L q \right),\label{dh7}\ee	
	\be +\fr{\Ga^{l Z^\prime}_{\al
			\beta}}{m_{Z^\prime}^2}\frac{gs_W^2}{3c_W\sqrt{1+2c_{2W}}}\left(\bar{l}_{\al}
	\ga^\mu P_L l_\beta  \right)
	\left(\eta^q \bar{q} \ga_\mu  P_R q \right),\label{dh8} \ee	
	\be 
	-\fr{1}{m_{Z^\prime}^2}\left(\fr{g(2+c_{2W})}{6c_W\sqrt{1+2c_{2W}}}\right)^2\left(\bar{q}
	\ga^\mu P_L q\right)\left(\bar{q}  \ga_\mu P_L
	q\right),\label{dh9}\ee	
	\be -\fr{1}{m_{Z^\prime}^2}\left(\frac{gs_W^2}{3c_W\sqrt{1+2c_{2W}}}\right)^2\left(\eta^q
	\bar{q}  \ga_\mu P_R q \right)
	\left(\eta^q \bar{q}  \ga_\mu P_R q \right).
	\label{eff1}\ee  
For convenience, we have relabeled the lepton eigenstates ($l'_i$) to be $l_\al$, which are determined by a Greece (letter) index $\al=1,2,3$ and should not be confused with a generation index, such that $e_\al=e,\mu,\tau$ and $\nu_\al= \nu_1,\nu_2,\nu_3$. A coefficient $\eta^q$ applies for the up-quarks as $\eta^u =2$ and the down-quarks as $\eta^d =-1$. 

The present constraints for the effective Lagrangian come from several processes. The first two terms (\ref{dh1}) and (\ref{dh2}) provide charged lepton flavor violating processes like $\mu \rightarrow 3e$, $\tau \rightarrow 3e$, $\tau \rightarrow 3 \mu$, $\tau \rightarrow 2e \mu$, and $\tau \rightarrow 2\mu e$. The next four terms (\ref{dh3}), (\ref{dh4}), (\ref{dh5}), and (\ref{dh6}) present wrong muon and tau decays as well as the nonstandard neutrino interactions that concern both constraints from oscillation and non-oscillation experiments. The last four terms (\ref{dh7}), (\ref{dh8}), (\ref{dh9}), and (\ref{eff1}) describe semileptonic $\tau\to \mu(e)$ decays and $\mu-e$ conversion in nuclei as well as the signals for new physics (dilepton, dijet, etc.) at low energy such as the Tevatron. Therefore, in the following, we consider the phenomenological aspect due to the presence of the interactions given above.
	
\section{\label{pheno}Phenomenology}
	
\subsection{Leptonic three-body decays}
  
To perform the analysis in the current section, we suppose that the major sources contributing to the lepton flavor violating processes come from the direct gauge interactions between the charged leptons and new massive gauge boson $Z^\prime$, while the others such as scalar contributions are considered to be small and neglected. The involved interactions, which are fully introduced in the previous section expressed in eqs. (\ref{dh1}-\ref{eff1}), have the amplitudes inversely proportional to the square of the new gauge boson mass, $m_{Z'}^2$, as well as dependent on the charged lepton mixing matrix $V_{eL}$. Moreover, the branching ratios or conversing ratio for $\mu\to e$ conversion in nuclei of the processes of interest are proportional to the square of the lepton-flavor violating effective interaction strengths, thus to be suppressed by the power fourth of the new gauge boson mass $m_{Z'}^4$.
	
The 3-by-3 unitary charged lepton mixing matrix $V_{eL}$ is undetermined, although $V^\dagger_{\nu L}V_{eL}$ is constrained by the neutrino oscillation data. In the current research, we parametrize $V_{eL}$ using three Euler angles $(\theta^\ell_{ij})$ and a phase 
	$(\delta^\ell)$, which is similar to the parametrizations of the CKM and PMNS matrices,  
	\begin{widetext}
\begin{equation}
\label{VeLMatrix}
V_{eL}=
\left(
\begin{array}{ccc}
 c_{12}c_{13} & s_{12}c_{13} & s_{13} e^{-i \delta^\ell} \\
 -c_{23}s_{12}-s_{13}s_{23}c_{12} e^{i \delta^\ell} & c_{23}c_{12}-s_{13}s_{23}s_{12}
 e^{i \delta^\ell} & s_{23}c_{13} \\
 s_{23}s_{12}-s_{13}c_{23}c_{12} e^{i \delta^\ell} & -s_{23}c_{12}-s_{13}c_{23}s_{12}
 e^{i \delta^\ell} & c_{23}c_{13}
\end{array}
\right).
\end{equation}
\end{widetext} 
Here, we use the notations $c_{ij}\equiv\cos\theta_{ij}^\ell$ and $s_{ij}\equiv\sin\theta_{ij}^\ell$, 
and the angles vary in the corresponding ranges, $\theta_{ij}^\ell=[0,\pi/2]$ and $\delta^\ell=[0,2 \pi]$. 
	 
\subsubsection{$\tau^+ \rightarrow \mu^+ \mu^+ \mu^-$, $\tau^+ \rightarrow e^+e^+e^-$ and $\mu^+ \rightarrow e^+e^+e^-$}
	
The lepton flavor violating processes of a lepton decaying into three identical lighter ones, which are usually called as the type I trilepton decays, have been considered before in various models and scenarios of elementary particle physics \cite{LFVOkda1,Ilakovac:1994kj, Blanke:2007db, Buras:2010cp, Dinh:2012bp, Dinh:2013vya}. On the experimental side, the temporary upper bound constraints on the branching ratios of those processes have also been obtained, such as \cite{Tanabashi:2018oca}
	\bea
	&& {\rm Br}(\mu\to 3e)< 1.0\times 10^{-12},\\
	&& {\rm Br}(\tau\to 3e)< 2.7\times 10^{-8},\\
	&& {\rm Br}(\tau\to 3\mu)< 2.1\times 10^{-8}.
	\eea
Thus, the upper bound on ${\rm Br}(\mu\to 3e)$ is about four orders more stringent than the corresponding channels of $\tau$ decays.
	
The lower bound on the new gauge boson mass $m_{Z'}$ comes from the LHC dilepton and dijet searches, which is roughly set to be larger than 4 TeV for the $Z'$ couplings analogous to the standard model $Z$ couplings \cite{Aaboud:2017buh,Aaboud:2017yvp}. However, in this model the $Z'$ mass takes a smaller bound, say 2.8 TeV, as shown below. Because the transferred momentum, whose maximal value is about the $\tau$ mass, is much smaller than $m_{Z'}$, the type I trilepton decay branching ratios, for instance $\tau^+ \rightarrow \mu^+\mu^+ \mu^-$, can be calculated with a high enough precision using the effective Lagrangian written as follows
	\bea
	\mathcal{L}^I= -\fr{4G_F}{\sqrt{2}}\left[ g_{LL}^I\left( \bar{\tau}\ga^\mu P_L
	\mu\right)\left( \bar{\mu}\ga_\mu P_L \mu\right)+ g_{LR}^I\left(
	\bar{\tau}\ga^\mu P_L \mu\right)\left( \bar{\mu}\ga_\mu P_R
	\mu\right)\right]+H.c.,
	\label{typeIa-LFV}\eea
	where \be g_{LL}^I=\fr{\sqrt{2}\Ga^{lZ^\prime}_{\tau \mu}\Ga^{lZ^\prime}_{\mu
			\mu}}{2G_Fm^2_{Z^\prime}},\hs
	g_{LR}^I=\left( \fr{\sqrt{2}\Ga^{lZ^\prime}_{\tau
			\mu}}{4G_Fm^2_{Z^\prime}}\right)\left(\fr{g s_W^2}{c_W\sqrt{1+2c_{2W}}}\right).\ee
The branching ratio for this process from eq. (\ref{typeIa-LFV}) can be found in \cite{LFVOkda1},
	\bea
	\mathrm{Br}(\tau^+ \rightarrow \mu^+ \mu^+
	\mu^-)=\left(|g^I_{LR}|^2+2|g^I_{LL}|^2\right)\mathrm{Br}(\tau^+ \rightarrow
	\bar{\nu}_\tau e^+\nu_e).
	\label{typeIb-LFV}\eea 
In the same way, the effective Lagrangians and branching ratios for $\tau^+ \rightarrow e^+e^+e^-$ and $\mu^+ \rightarrow e^+e^+e^-$ channels can be derived. That said, their expressions could be obtained from (\ref{typeIa-LFV}) and (\ref{typeIb-LFV}) by replacing $\mu$ by $e$ for $\tau\to 3e$ and ($\tau$, $\mu$) by ($\mu$, $e$) for $\mu\to 3e$, respectively.
	
Theoretically, to set the upper bound on ${\rm Br}(\ell\to 3\ell')$ in the flipped 3-3-1 model, we vary $\theta_{ij}^\ell$ in $[0,\pi/2]$ and 
	$\delta^\ell$ in $[0,2\pi]$. The result is:
	\begin{eqnarray}
	0\leq {\rm Br}(\mu\to 3e)\leq 4.4272\times 10^{-5}\left(\frac{1\ {\rm TeV}}{m_{Z'}}\right)^4,\\
	0\leq {\rm Br}(\tau\to 3e)\leq 7.8892\times 10^{-6}\left(\frac{1\ {\rm TeV}}{m_{Z'}}\right)^4,\\
	0\leq {\rm Br}(\tau\to 3\mu)\leq 7.6989\times 10^{-6}\left(\frac{1\ {\rm TeV}}{m_{Z'}}\right)^4.
	\end{eqnarray}
Here the minimal (maximal) values frequently occur at many relevant values of $(\theta_{ij}^\ell, \delta^\ell)$, for example, at
$(\sin\theta_{12}^\ell,\sin\theta_{13}^\ell,\sin\theta_{23}^\ell,\delta^\ell)$:
\begin{eqnarray}
0.7074,~0.0000,~0.4336,~1.04\pi~~~(0.1005,~1.0000,~0.8601,~1.70\pi),\\
0.0000,~0.7071,~0.2397,~0.45\pi~~~(0.8498,~0.0000,~0.0464,~0.99\pi),\\
1.0000,~0.7072,~0.5033,~0.39\pi~~~(0.0000,~0.0000,~0.8082,~0.38\pi),
\end{eqnarray}
according to ${\rm Br}(\mu\to 3e)$, ${\rm Br}(\tau\to 3e)$, and ${\rm Br}(\tau\to 3\mu)$, respectively. Indeed, the minimum is easily seen to be at the points that $\Ga^{lZ^\prime}_{\mu e}$, 
$\Ga^{lZ^\prime}_{\tau e}$, and $\Ga^{lZ^\prime}_{\tau \mu}$ vanish, respectively.
 
%%%%%%%%%%%%%%%%%%%%%%%%%%%%%%%
\begin{figure}[t]
\begin{center}
\begin{tabular}{cc}
\includegraphics[width=7.5cm,height=6.5cm]{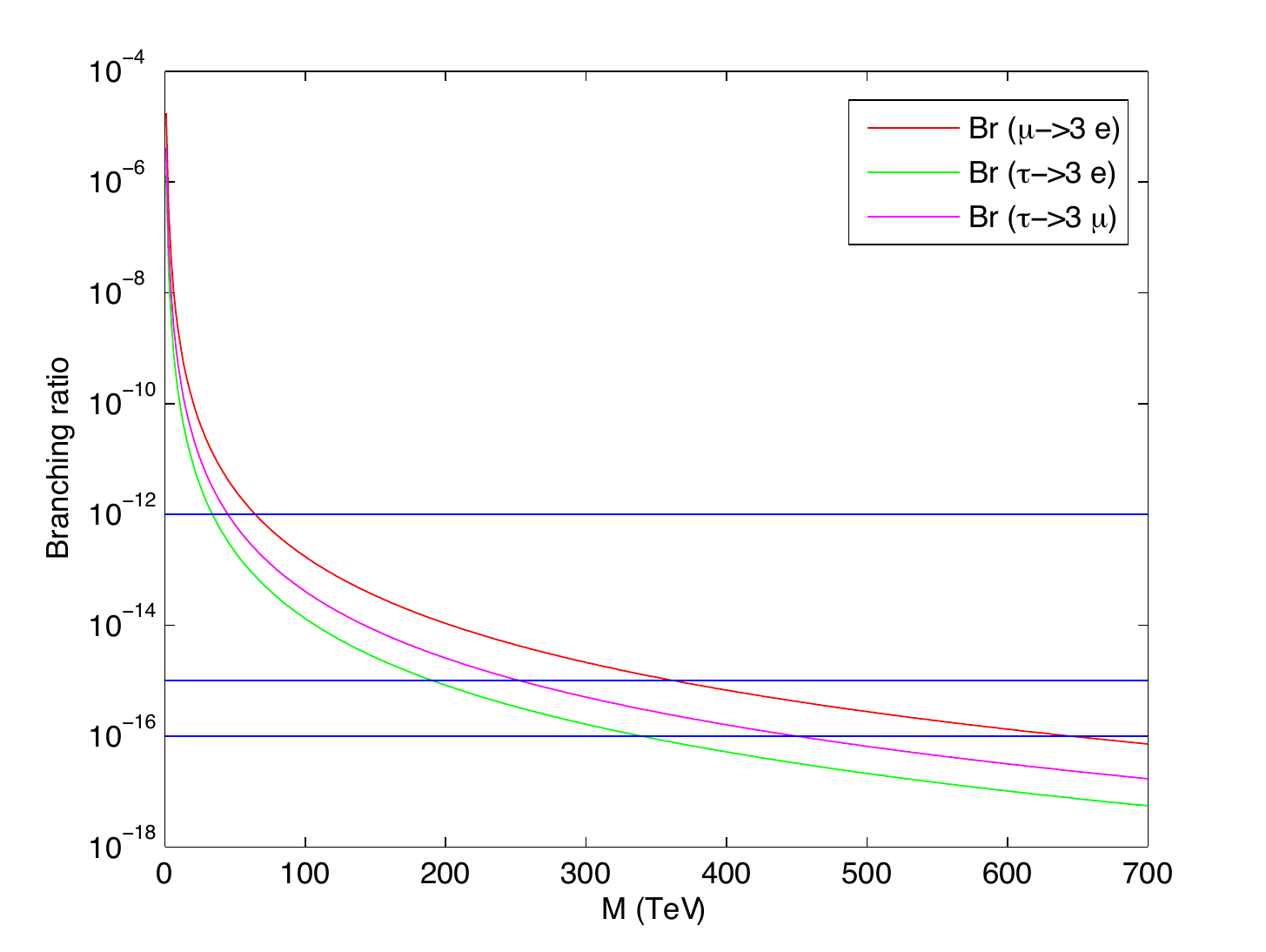} &
\includegraphics[width=7.5cm,height=6.5cm]{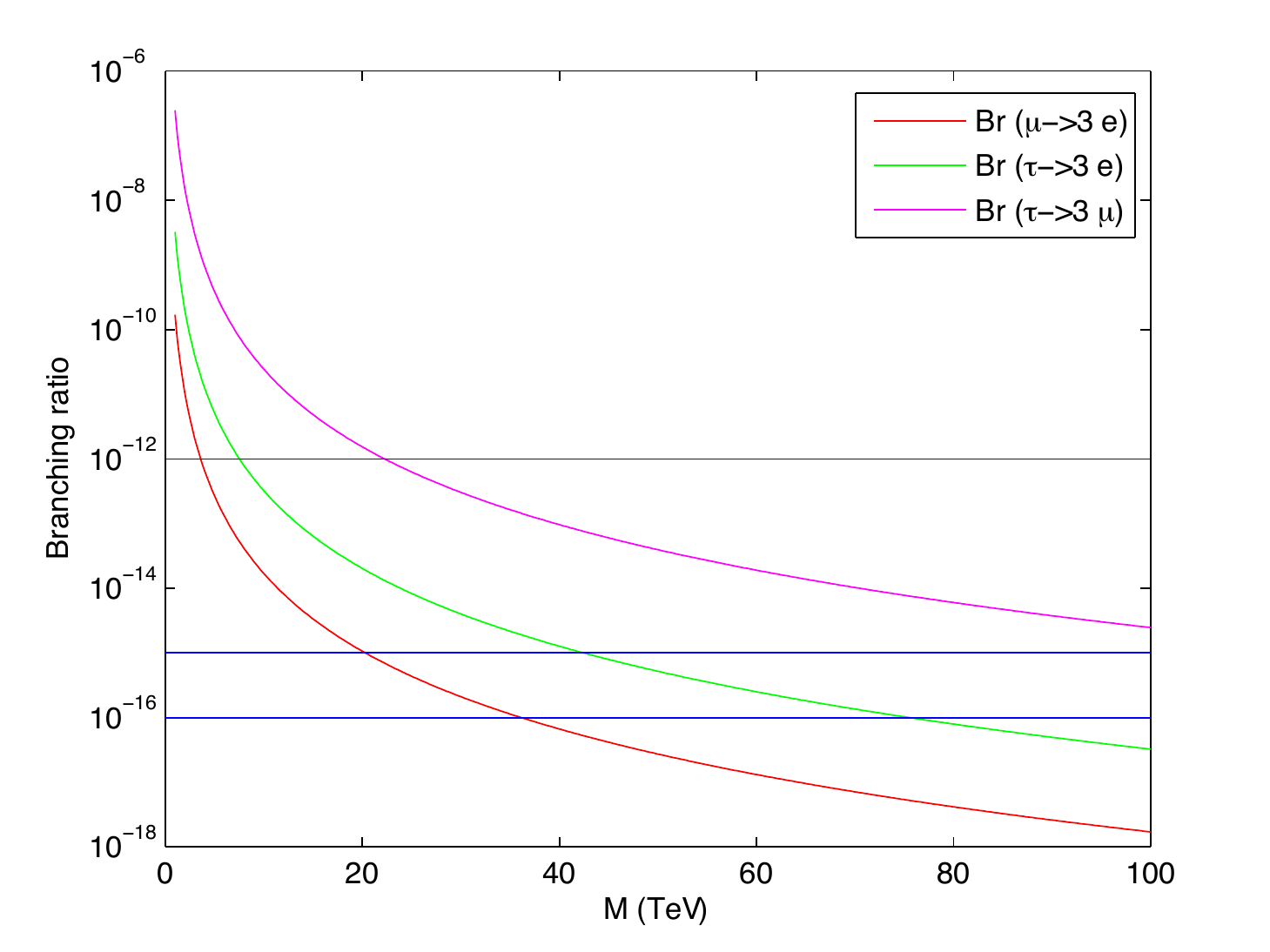}
\end{tabular}
 \caption{Branching ratios ${\rm Br}(\mu\to 3e)$, ${\rm Br}(\tau\to 3e)$, and ${\rm Br}(\tau\to 3\mu$) as 
functions of the new neutral gauge boson mass $m_{Z'}\equiv M$, respectively. The three blue
horizon-lines correspond to the currently experimental upper bound and expected sensitivities of the PSI and PSI upgraded experiments, namely ${\rm Br}(\mu\to 3e) = 10^{-12}$, $10^{-15}$, and $10^{-16}$, respectively. The left-panel is created for $\theta_{12}^\ell=\pi/3$, $\theta_{13}^\ell=\pi/6$,
$\theta_{23}^\ell=\pi/4$, and $\de^\ell=0$, whereas the right-panel is produced according to $\sin\theta_{12}^\ell=0.9936$, $\sin\theta_{13}^\ell=0.9953$, $\sin\theta_{23}^\ell=0.2324$, and $\delta^\ell=1.10\pi$.}
\label{FigMu3e}
\end{center}
\end{figure}

In Fig. \ref{FigMu3e}, we illustrate the dependence of the branching ratio ${\rm Br}(\ell\to 3\ell')$ as
a single variable function of the new gauge boson mass $m_{Z'}$ for $\theta_{12}^\ell=\pi/3$, $\theta_{13}^\ell=\pi/6$, 
$\theta_{23}^\ell=\pi/4$, and $\delta^\ell=0$ in the left panel, and for $\sin\theta_{12}^\ell=0.9936$, $\sin\theta_{13}^\ell=0.9953$, $\sin\theta_{23}^\ell=0.2324$, $\delta^\ell=1.10\pi$ in the right panel. The currently experimental 
upper bounds on the $\tau$ decay channels are
omitted from the figure because of their much less stringency comparing to those of $\mu$. As shown in Fig. \ref{FigMu3e}, the branching ratio lines 
decreasing as $m_{Z'}$ increasing are consistent with the fact that they are inversely proportional to 
$m_{Z'}^4$, aforementioned. Using the results shown in Fig. \ref{FigMu3e}, one obtains lower limits $m_{Z'}\geq 65.3$, $362.1$, and $652.4$ TeV for the left panel and $m_{Z'}\geq 3.8$, $20.6$, and $36.5$ TeV for the right panel, corresponding to the current upper bound and future sensitivities of the PSI and PSI upgraded experiments, respectively. The latter values quite agree with the current collision bounds on $Z'$ mass. 

\subsubsection{$\tau^+ \rightarrow \mu^+ e^+ e^-$ and $\tau^+ \rightarrow e^+\mu^+\mu^-$}
	
Similar to the decays $\ell\to 3\ell'$ considered in the previous section, when the new gauge boson mass $m_{Z'}$ is at order TeV or larger, the type II three leptonic decay, e.g. $\tau^+ \rightarrow \mu^+e^+e^-$, can be well described by an effective Lagrangian, which takes the form
	\bea
	\nonumber
	\mathcal{L}^{II} &=& -\fr{4G_F}{\sqrt{2}}\left[ g_{LL}^{II}\left( \bar{\tau}\ga^\mu
	P_L \mu\right)\left( \bar{e}\ga_\mu P_L e\right)
	+g_{LL}^{X}\left( \bar{\tau}\ga^\mu P_L e\right)\left( \bar{e}\ga_\mu P_L \mu\right)\right.\\ 
	&& \left.+ g_{LR}^{II}\left(\bar{\tau}\ga^\mu P_L \mu\right)\left( \bar{e}\ga_\mu P_R e\right)\right]+H.c.,
	\eea where 
	\bea g_{LL}^{II}=\fr{\sqrt{2}\Ga^{lZ^\prime}_{\tau \mu}\Ga^{lZ^\prime}_{ee
	}}{2G_Fm^2_{Z^\prime}},\hs g_{LL}^{X}=\fr{\sqrt{2}\Ga^{lZ^\prime}_{\tau e}\Ga^{lZ^\prime}_{e\mu
	}}{2G_Fm^2_{Z^\prime}},\hs
	g_{LR}^{II}=\left( \fr{\sqrt{2}\Ga^{lZ^\prime}_{\tau
			\mu}}{4G_Fm^2_{Z^\prime}}\right)\left(\fr{g s_W^2}{c_W\sqrt{1+2c_{2W}}}\right).\eea
Then the branching ratio for $\tau^+ \rightarrow \mu^+e^+ e^-$ decay could be 
	expressed as \cite{LFVOkda1},
	\bea
	\mathrm{Br}(\tau^+ \rightarrow \mu^+
	e^+e^-)=\left(|g^{II}_{LR}|^2+|g^{II}_{LL}|^2+|g^{X}_{LL}|^2\right)\mathrm{Br}(\tau^+ \rightarrow
	\bar{\nu}_\tau e^+\nu_e).
	\eea
With the aid of the interchange symmetry $\mu \leftrightarrow e$, the expression for the branching ratio of the decay $\tau^+ \rightarrow e^+\mu^+\mu^-$ could be easily obtained from the above formula by appropriately replacing $(\Ga^{lZ^\prime}_{\tau \mu}, \Ga^{lZ^\prime}_{ee})$ by  $(\Ga^{lZ^\prime}_{\tau e}, \Ga^{lZ^\prime}_{\mu \mu}$).
%%%%%%%%%%%%%%%%%%%%%%%
\begin{figure}[t]
\begin{center}
\includegraphics[width=11cm]{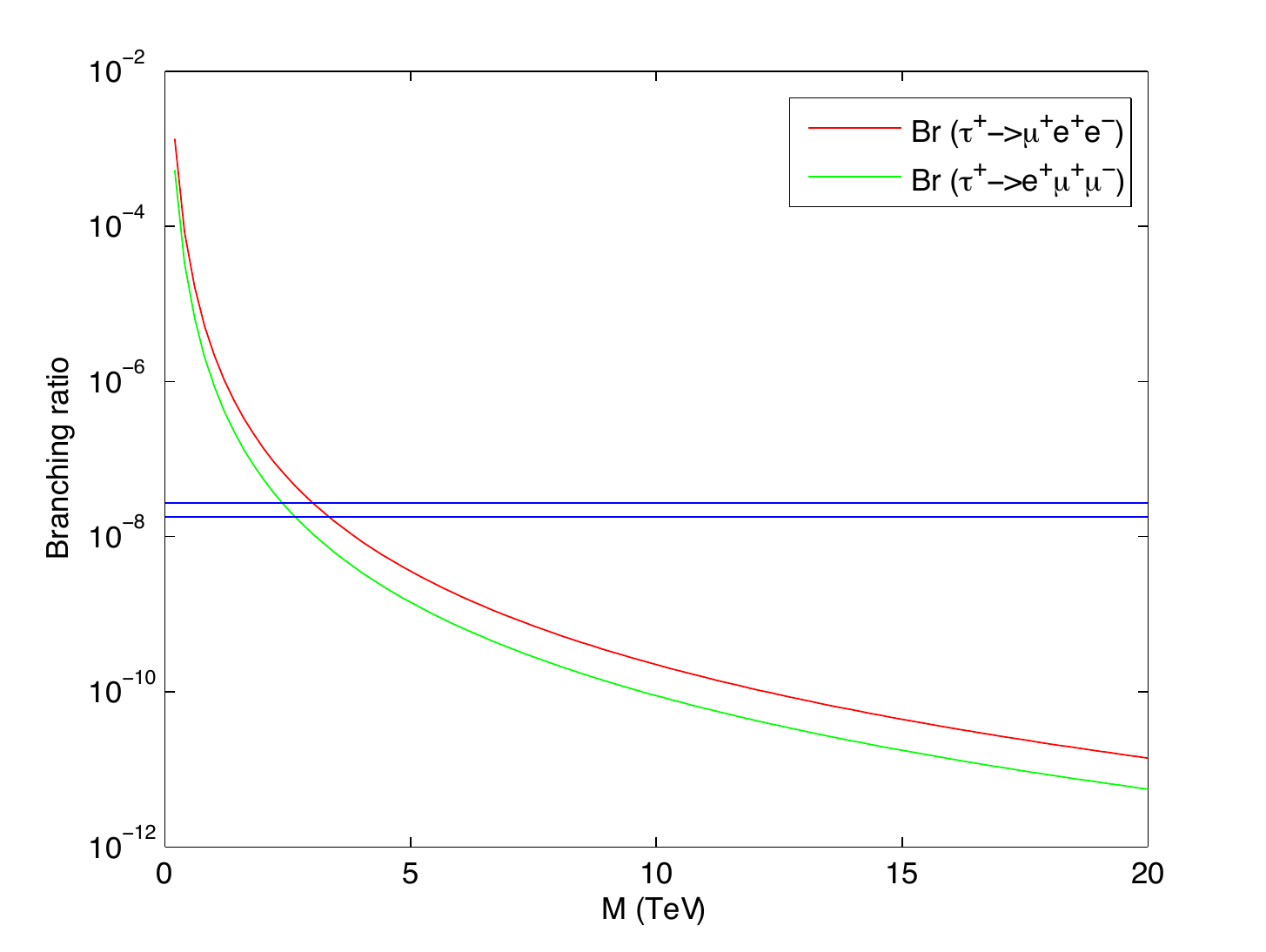}
   \caption{Dependence of the branching ratios $\mathrm{Br}(\tau\to e\mu\mu)$ and 
   $\mathrm{Br}(\tau\to \mu e e)$ on the new neutral gauge boson mass, $m_{Z'}\equiv M$. Here, the blue 
   lines correspond to the current upper limits $\mathrm{Br}(\tau\to e\mu\mu)\leq 2.7\times 10^{-8}$ and 
   $\mathrm{Br}(\tau\to \mu ee)\leq 1.8\times 10^{-8}$, respectively.}
\label{TauMuee}
 \end{center}
\end{figure}
%%%%%%%%%%%%%%%%%%%%%%%%%%%%%%%% 

Taking into account the same set of values of $\theta_{ij}^\ell$ and $\delta^\ell$ as in the previous section, we depict in Fig.~\ref{TauMuee} the behaviors of the branching ratios, $\tau\to e\mu\mu$ and $\tau\to \mu e e$, according to the variation of the new neutral gauge boson mass $m_{Z'}$ (green and red lines), respectively. In this figure, the current upper bounds $\mathrm{Br}(\tau\to e\mu\mu)\leq 2.7\times 10^{-8}$ and $\mathrm{Br}(\tau\to \mu ee)\leq 1.8\times 10^{-8}$ are also shown as blue lines \cite{Tanabashi:2018oca}. 
The lower limits for $m_{Z'}$ obtained from Fig. \ref{TauMuee} responsible for these two processes are roughly $3$ TeV. This agrees with the limit from $\mu \to 3e$ in the case of Fig. \ref{FigMu3e} right panel, but is about 20 times less stringent than the limit from $\mu\to 3e$ in the case of Fig. \ref{FigMu3e} left panel. Particularly, using the current lower bound from $\mu\to 3e$ in the latter case, i.e. $m_{Z'}\geq 65.3$ GeV at 
   $\theta_{12}^\ell=\pi/3$, $\theta_{13}^\ell=\pi/6$, $\theta_{23}^\ell=\pi/4$, and $\delta^\ell=0$, the precision calculation shows that 
   \begin{eqnarray}
	{\rm Br}(\tau\to\mu ee)\leq 4.88\times 10^{-14},\\
	{\rm Br}(\tau\to e\mu\mu)\leq 1.31\times 10^{-13}.
\end{eqnarray}

One the other hand, with the same strategy as in the previous case when varying $\theta_{ij}^\ell$ in $[0,\pi/2]$ and $\delta^\ell$ in $[0,2\pi]$,
the bounds derived are 
\begin{eqnarray}
	0\leq {\rm Br}(\tau\to\mu ee)\leq 4.4562\times 10^{-6}\left(\frac{1\ {\rm TeV}}{m_{Z'}}\right)^4,\\
	0\leq {\rm Br}(\tau\to e\mu\mu)\leq 4.4562\times 10^{-6}\left(\frac{1\ {\rm TeV}}{m_{Z'}}\right)^4.
\end{eqnarray}

\subsubsection{$\tau^+ \rightarrow \mu^+ \mu^+ e^-$ and $\tau^+ \rightarrow e^+e^+\mu^-$}
	
In this part of search, we consider the type III leptonic decay modes in which lepton flavors are violated by three units, such as $\tau^+ \rightarrow \mu^+ \mu^+ e^-$. In the flipped 3-3-1 model, the	 process is dominantly contributed by an unique effective interaction, which is expressed as 
	\bea
	\mathcal{L}^{III}=-\fr{4G_F}{\sqrt{2}}\left[g_{LL}^{III}\left(\bar{\tau}\ga^\mu
	P_L \mu\right) \left(\bar{e}\ga_\mu P_L \mu \right) \right]+H.c.,
	\label{LFV3a} \eea 
where \be g_{LL}^{III}=\fr{\sqrt{2}\Ga^{lZ^\prime}_{\tau \mu}\Ga^{lZ^\prime}_{e\mu
	}}{2G_Fm^2_{Z^\prime}}.\ee
	
Then, the branching ratio for $\tau^+ \rightarrow \mu^+ \mu^+
	e^-$ could be easily written down by generalizing the results given in \cite{LFVOkda1} as
	\bea
	\mathrm{Br}(\tau^+ \rightarrow \mu^+ \mu^+ e^-) =2|g_{LL}^{III}|^2\mathrm{Br}(\tau^+ \rightarrow
	\bar{\nu}_\tau e^+\nu_e).
	\label{LFV3aa}
	\eea
On the other hand, the effective Lagrangian and branching ratio for $\mathrm{Br}(\tau^+ \rightarrow e^+e^+ \mu^-)$ can be achieved from (\ref{LFV3a}) and (\ref{LFV3aa}) simply by interchanging the fields $\mu\leftrightarrow e$.
	
\begin{figure}[t]
\begin{center}
\includegraphics[width=11cm]{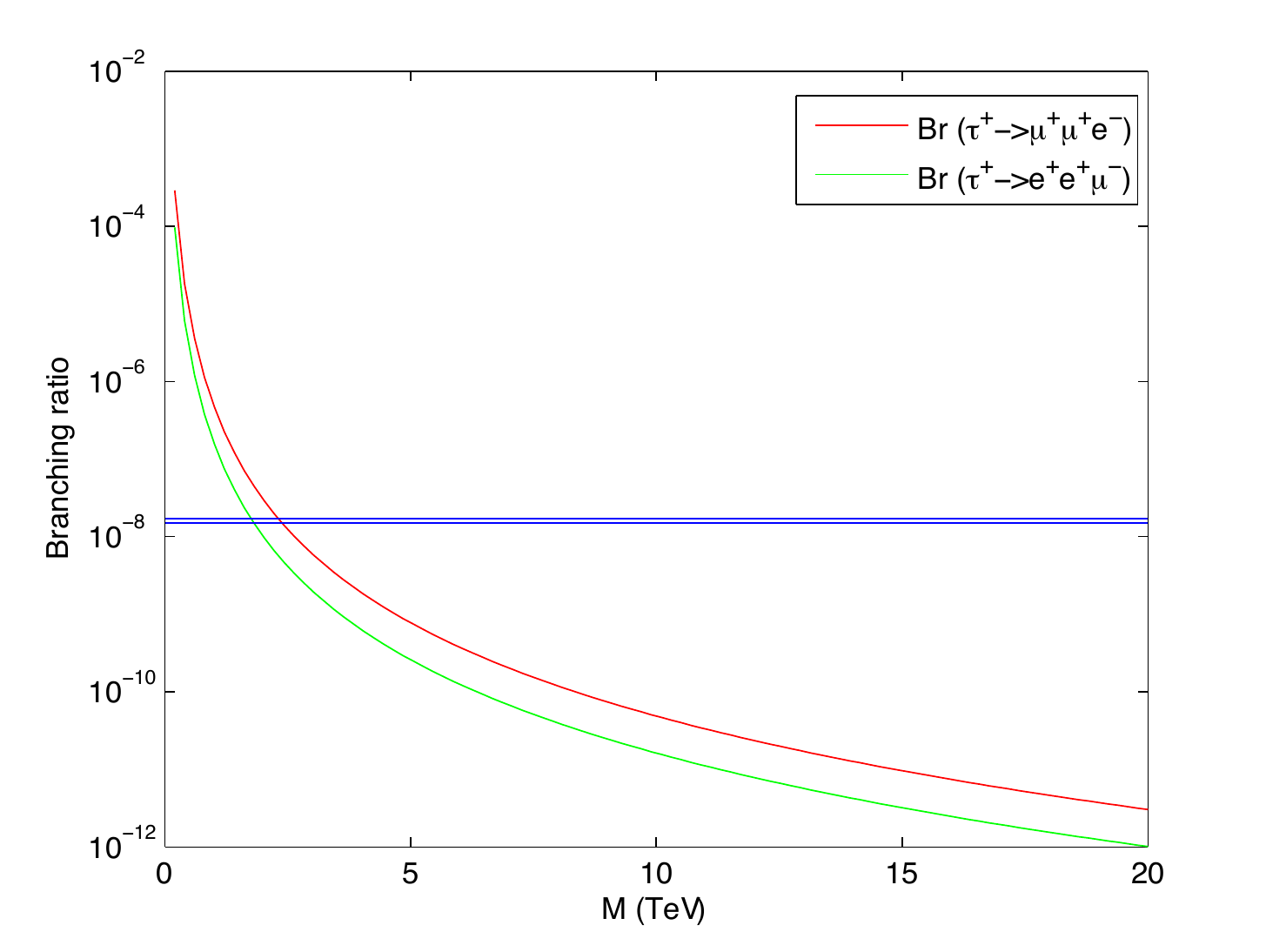}
 \caption{Dependence of the branching ratios $\mathrm{Br}(\tau\to \mu\mu e)$ and 
   $\mathrm{Br}(\tau\to e\mu\mu)$ on the new neutral gauge boson mass $m_{Z'}\equiv M$. The blue 
   lines correspond to the current upper limits $\mathrm{Br}(\tau\to \mu\mu e)\leq 1.7\times 10^{-8}$ and 
   $\mathrm{Br}(\tau\to ee\mu)\leq 1.5\times 10^{-8}$, which are almost coincided \cite{Tanabashi:2018oca}.}
\label{TauMuMue}
 \end{center}
\end{figure}

Currently, the experimental constraints on the branching ratios of the decays $\tau^+ \to\mu^+ \mu^+ e^-$ and $\tau^+ \to e^+e^+\mu^-$ are, indeed, very weak. Therefore, the corresponding lower limits on the new neutral gauge boson mass $m_{Z'}$ given by these two channels are less substantial than those obtained from the channels considered above, especially for the $\mu\to 3e$ decay. To be concrete, the behaviors of the $\tau^+ \to\mu^+ \mu^+ e^-$ and $\tau^+ \to
	e^+e^+\mu^-$ branching ratios can be found in Fig. \ref{TauMuMue}. Using the same trick as in the previous case, we obtain the theoretical upper bounds as follows
	%%%%%%%%%%%%%%%%%%%%%%%
%%%%%%%%%%%%%%%%%%%%%%%%%%%%%%%% 
\bea
	&&{\rm Br}(\tau\to\mu\mu e)\leq 3.20\times 10^{-14},\\
	&&{\rm Br}(\tau\to ee\mu)\leq 8.17\times 10^{-15},
\eea
for $\theta_{12}^\ell=\pi/3$, $\theta_{13}^\ell=\pi/6$, $\theta_{23}^\ell=\pi/4$, and $\delta^\ell=0$. Hence, these bounds are about six orders below the sensitivities of the current experiments.

Last, but not least, varying the mixing angles and phase in the allowed regimes, the specified ranges for
the $\tau^+ \to\mu^+ \mu^+ e^-$ and $\tau^+ \to e^+e^+\mu^-$ branching ratios are found as
%%%%%%%%%%%%%%%%%%%%%%%%%%%%%%%% 
	\begin{eqnarray}
	&& 0\leq {\rm Br}(\tau\to\mu\mu e)\leq 5.1155\times 10^{-7}\left(\frac{1\ {\rm TeV}}{m_{Z'}}\right)^4,\\
	&& 0\leq {\rm Br}(\tau\to ee\mu)\leq 5.1155\times 10^{-7}\left(\frac{1\ {\rm TeV}}{m_{Z'}}\right)^4.
\end{eqnarray}

\subsubsection{Comment on wrong $\mu$ and $\tau$ decays}

It is not hard to point out that the wrong muon and tau decays, e.g. $\mu \to e \nu_e\bar{\nu}_\mu$ and $\tau\to \mu \nu_\mu \bar{\nu}_\tau$, take the same rate as of those in the previous section, respectively. Hence, such decays are far below the experimental limits $\mathrm{Br}\sim 0.1$ \cite{Tanabashi:2018oca}.

\subsection{Semileptonic $\tau \rightarrow \mu$ and $\tau \rightarrow e$ decays}
		
The next topic we discuss in this paper is the semileptonic decays of $\tau$, say \be \mathrm{Br}(\tau^+ \rightarrow \ell^+P)\ \mathrm{and}\ \mathrm{Br}(\tau^+ \rightarrow \ell^+V),\ee in which $\ell=e,\mu$ and $P,V$ stand for neutral pseudoscalar and vector mesons, respectively. These decay channels have been studied formerly in the other models as well as model-independent scenarios \cite{Ilakovac:1995km,LFVOkda1,He:2019iqf}. In the present model, these processes happen dominantly through the exchange of the new neutral gauge boson with lepton-flavor violating interactions. On the 
experimental side, the following upper bounds have been obtained \cite{Tanabashi:2018oca}:
\begin{align} \label{limits}
& {\rm Br}(\tau^+\to e^+\pi^0) ~<\, 8.0\times10^{-8} \,, &
& {\rm Br}(\tau^+\to\mu^+\pi^0) ~<\, 1.1\times10^{-7} \,,
\nonumber \\
& {\rm Br}(\tau^+\to  e^+\eta^0) \,<\, 9.2\times10^{-8} \,, &
& {\rm Br}(\tau^+\to\mu^+\eta^0) \,<\, 6.5\times10^{-8} \,,
\nonumber \\
& {\rm Br}(\tau^+\to  e^+\eta'^{0}) \,<\, 1.6\times10^{-7} \,, &
& {\rm Br}(\tau^+\to\mu^+\eta'^0) \,<\, 1.3\times10^{-7} \,,\\
& {\rm Br}(\tau^+\to e^+\rho^0) ~<\, 1.8\times10^{-8} \,, &
& {\rm Br}(\tau^+\to\mu^+\rho^0) ~<\, 1.2\times10^{-8} \,,
\nonumber \\
& {\rm Br}(\tau^+\to  e^+\omega^0) \,<\, 4.8\times10^{-8} \,, &
& {\rm Br}(\tau^+\to\mu^+\omega^0) \,<\, 4.7\times10^{-8} \,,
\nonumber \\
& {\rm Br}(\tau^+\to  e^+\phi^{0}) \,<\, 3.1\times10^{-8} \,, &
& {\rm Br}(\tau^+\to\mu^+\phi^0) \,<\, 8.4\times10^{-8}.\nonumber
\end{align}
The near future and planned experiments such as LCHb \cite{Cerri:2018ypt}, BESIII \cite{Li:2012vk}, 
Belle II \cite{Kou:2018nap}, and COMET \cite{Calibbi:2017uvl} will improve upon 
these present limits. 

As already supplied in (\ref{dh7}) and (\ref{dh8}), the effective Lagrangian for the semileptonic decays of the 
$\tau\rightarrow \mu(e)$ types can be rewritten in the form
	\bea
	\mathcal{L}_{Had}^I= -\fr{4G_F}{\sqrt{2}}\left[
	h_{LL}^{q\ell}\left(\bar{\tau}\ga^\mu P_L \ell \right)\left(\bar{q}\ga_\mu P_L
	q\right)+ h_{LR}^{q\ell}\left(\bar{\tau}\ga^\mu P_L \ell \right)\left(\bar{q}\ga_\mu
	P_R q\right)\right]+H.c.,
	\label{Semi1}\eea
where $\ell=\mu$ or $e$ and the couplings are 
\be h_{LL}^{q\ell}=\fr{g\sqrt{2}(2+c_{2W})\Ga_{\tau
			\ell}^{lZ^\prime}}{24G_Fm_{Z^\prime}^2c_W\sqrt{1+2c_{2W}}},\hs  
	h_{LR}^{q\ell}=\fr{g\sqrt{2}s_W^2 \Ga^{l Z^\prime}_{\tau
			\ell}\eta_q}{12G_Fm_{Z^\prime}^2c_W\sqrt{1+2c_{2W}}}.\ee
	
The branching ratio for the $\tau^+ \rightarrow \ell^+P $ decay, where $P$ denotes a neutral pseudoscalar meson as $\pi^0$ or $\eta^0$ or $\eta'^0$, can be derived from
	(\ref{Semi1}) as follows \cite{LFVOkda1}
	\bea
	\mathrm{Br}(\tau^+ \rightarrow\ell^+ P)=\tau_\tau \fr{G_F^2
		m_\tau^3}{4\pi}\left(1-\fr{m_P^2}{m_\tau^2}\right)|C_{eff}^{\ell}|^2,
	\eea
where $\tau_\tau$ is the tau's lifetime and 
	\be C_{eff}^\ell=\fr{f_{\eta,\eta^\prime}^q}{2\sqrt{2}}\left(h_{LR}^{u\ell}+h_{LR}^{d\ell}-h_{LL}^{u\ell}-h_{LL}^{d\ell}\right)+
	\fr{f^s_{\eta,\eta^\prime}}{2}\left(h_{LR}^{d\ell}-h_{LL}^{d\ell}\right),\ee for $P=\eta^0$ or $P=\eta'^0$, and  
	\be C_{eff}^\ell=\fr{f_\pi}{2\sqrt{2}}\left(h_{LR}^{u\ell}-h_{LR}^{d\ell}-h_{LL}^{u\ell}+h_{LL}^{d\ell}\right),\ee for $P=\pi^0$. Here the decay constant $f_\pi=135$ MeV, while $f^q_{\eta,\eta^\prime}$ and $f_{\eta,\eta^\prime}^s$ are defined as
	\begin{equation}
	\left(\begin{array}{cc}
		f_\eta^q & f_\eta^s \\
		f_\eta'^q & f_\eta'^s \\
	\end{array}\right)=\left(\begin{array}{cc}
		f_q\cos \phi_\eta  & -f_s\sin \phi_\eta \\
		f_q\sin \phi_\eta  & f_s\cos \phi_\eta \\
	\end{array}\right),
	\end{equation}
with $f_q\simeq 1.07 f_\pi,~f_s\simeq 1.34 f_\pi$ and $\phi_\eta\simeq 0.2183\pi$.
	
Based on the current experimental limits, our numerical calculation will show that the branching ratios of semileptonic $\tau$ decays into the pseudoscalar mesons are actually small to impose meaningful constraints on the relevant parameters in the flipped 3-3-1 model. For concreteness, the results of varying $\theta_{ij}^\ell$ and $\delta^\ell$ in the allowed ranges lead to
\begin{eqnarray}
    \nonumber
	0\leq {\rm Br}(\tau\to e(\mu)\pi)\leq 9.6517\times 10^{-8}\left(\frac{1\ {\rm TeV}}{m_{Z'}}\right)^4,\\
	\label{BRTu2LP}
	0\leq {\rm Br}(\tau\to e(\mu)\eta)\leq 2.3196\times 10^{-6}\left(\frac{1\ {\rm TeV}}{m_{Z'}}\right)^4,\\
	\nonumber
	0\leq {\rm Br}(\tau\to e(\mu)\eta')\leq 2.0475\times 10^{-6}\left(\frac{1\ {\rm TeV}}{m_{Z'}}\right)^4.
\end{eqnarray}

Let us note that, although the maximal values of the branching ratios obtained above do not depend on the lepton types, namely 
${\rm Max}[{\rm Br}(\tau\to e P)]={\rm Max}[{\rm Br}(\tau\to \mu P)]$ for any given $P$, they reach, however, the maximal values for different sets of lepton mixing parameters. For instance, ${\rm Br}(\tau\to e \pi)$ has 
the maximal value at $\sin\theta_{12}^\ell=0.0000$, $\sin\theta_{13}^\ell=0.7072$, $\sin\theta_{23}^\ell=0.7295$, 
and $\delta^\ell=1.13\pi$, whereas ${\rm Br}(\tau\to e \eta)$ gets the maximal value when $\sin\theta_{12}^\ell=1.0000$, $\sin\theta_{13}^\ell=0.7073$, $\sin\theta_{23}^\ell=0.8195$, and $\delta^\ell=1.92\pi$.
%
%%%%%%%%%%%%%%%%%%%%%%%
\begin{figure}[t]
\begin{center}
\includegraphics[width=11cm]{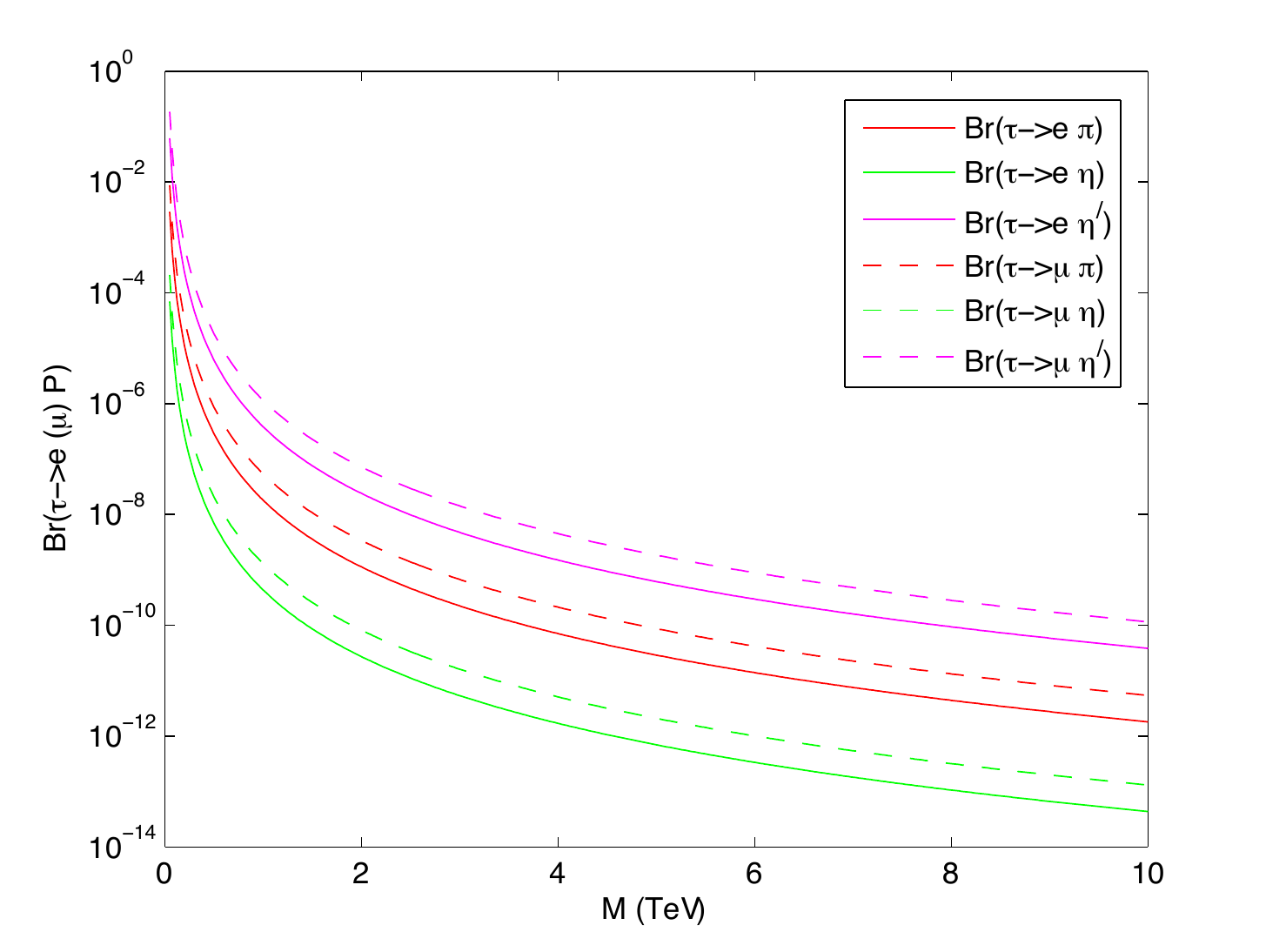}
 \caption{Dependence of the branching ratios $\mathrm{Br}(\tau^+\to \ell^+ P)$ on the new neutral gauge boson 
 mass $m_{Z'}\equiv M$, where $\ell=e,\mu$ and $P=\pi,\eta,\eta'$. Here, lepton mixing angles and 
 phase $\theta_{12}^\ell=\pi/3$, $\theta_{13}^\ell=\pi/6$, $\theta_{23}^\ell=\pi/4$, and $\delta^\ell=0$ have been used.}
\label{SemiTauP}
 \end{center}
\end{figure}
%%%%%%%%%%%%%%%%%%%%%%%%%%%%%%%% 

The theoretical maximal values given in (\ref{BRTu2LP}) are roundly the same orders as the current 
experimental limits [cf. (\ref{limits})]. Thus the constraints on the new neutral gauge boson mass 
derived from these channels have much appealing, validating the model. This can be more visually seen in 
Fig. \ref{SemiTauP}, where we show the dependence of the branching ratios $\mathrm{Br}(\tau^+\to \ell^+ P)$ 
on the new neutral gauge boson mass $m_{Z'}\equiv M$ for $\ell=e,\mu$ and $P=\pi,\eta,\eta'$.  
Since the current upper bounds on the branching ratios are around $10^{-7}$, the lower limit obtained for the new neutral gauge boson mass $m_{Z'}$ is about $3$ TeV, which is the same limit set by  
the searches of the LHC dilepton and dijet signals.
	
Similar conclusions are also obtained for the case of $\tau^+ \rightarrow \ell^+ V$ decay, where $V$ is taken as a neutral vector meson $\rho$ or $\omega$ or  $\phi$. The branching ratios for these processes can be derived from the Lagrangian (\ref{Semi1}) as \cite{LFVOkda1}
	\bea
	Br(\tau^+ \rightarrow \ell^+V)=\tau_\tau \fr{G_F^2
		m_\tau^3}{\pi}\left(1-\fr{m_V^2}{m_\tau^2}
	\right)^2\left(\fr{m_\tau^2+2m_V^2}{4m_V^2}\right)|G_{LV}^\ell|^2,
	\eea
	where $G_{LV}^\ell$ is an effective coupling that takes one of the following forms,
	\begin{itemize}
		\item $G_{LV}^\ell=\fr{f_\rho m_\rho}{2\sqrt{2}
			m_\tau}\left(h_{LL}^{u\ell}-h_{LL}^{d\ell}+h_{LR}^{u\ell}-h_{LR}^{d\ell}\right)$ for $V=\rho$.
		\item $G_{LV}^\ell=\fr{f_\omega
			m_\omega}{2\sqrt{2}m_\tau}\left(h_{LL}^{u\ell}+h_{LL}^{d\ell}+h_{LR}^{u\ell}+h_{LR}^{d\ell}\right)$ for
		$V=\omega$.
		\item $G_{LV}^\ell=-\fr{f_\phi
			m_\phi}{2m_\tau}\left(h_{LL}^{s\ell}+h_{LR}^{s\ell}\right)$ for $V=\phi$.
	\end{itemize}
	The coefficients $f_\rho=221$ MeV, $f_\omega=196$ MeV, and $f_\phi=228$ MeV are the form factors of the neutral vector mesons  $\rho$, $\omega$, and $\phi$, respectively.
	
\begin{figure}[t]
\begin{center}
\includegraphics[width=11cm]{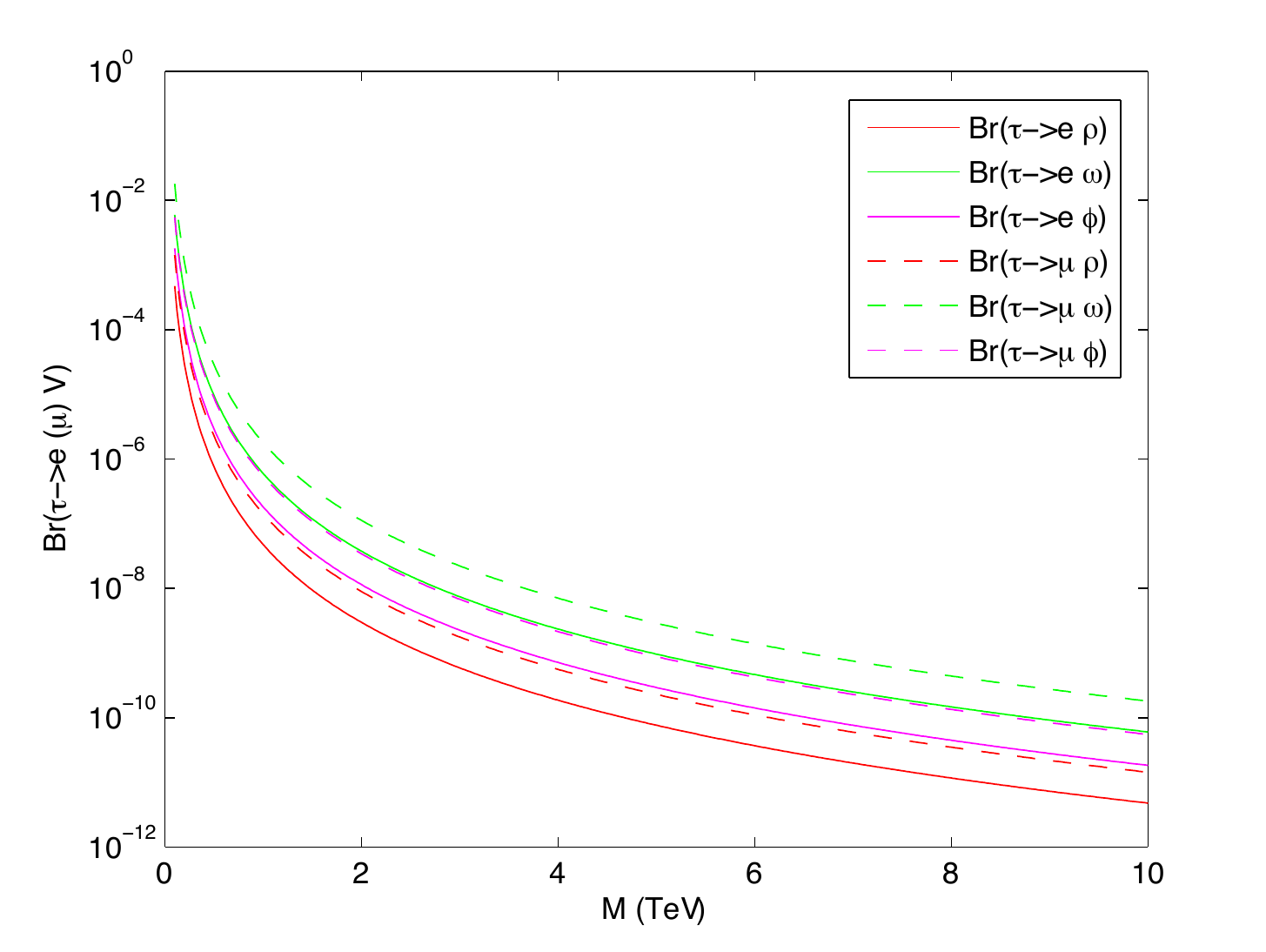}
 \caption{Dependence of the branching ratios $\mathrm{Br}(\tau^+\to \ell^+ V)$ on the new neutral gauge boson 
 mass $m_{Z'}\equiv M$, where $\ell=e,\mu$ and $P=\rho,\omega,\phi$. To produce the figure, lepton mixing angles and 
 phase $\theta_{12}^\ell=\pi/3$, $\theta_{13}^\ell=\pi/6$, $\theta_{23}^\ell=\pi/4$, and $\delta^\ell=0$ have been taken.}
\label{SemiTauV}
 \end{center}
\end{figure}

The identified ranges of the branching ratios when varying the parameters $\theta_{ij}^\ell$ and $\delta^\ell$ in their domains are
\begin{eqnarray}
	0\leq {\rm Br}(\tau\to e(\mu)\rho)\leq 2.5390\times 10^{-7}\left(\frac{1\ {\rm TeV}}{m_{Z'}}\right)^4,\\
	0\leq {\rm Br}(\tau\to e(\mu)\omega)\leq 3.2013\times 10^{-6}\left(\frac{1\ {\rm TeV}}{m_{Z'}}\right)^4,\\
	0\leq {\rm Br}(\tau\to e(\mu)\phi)\leq 9.4756\times 10^{-7}\left(\frac{1\ {\rm TeV}}{m_{Z'}}\right)^4.
\end{eqnarray}
The detail behaviors of the branching ratios $\mathrm{Br}(\tau^+\to \ell^+ V)$ are depicted in Fig. \ref{SemiTauV}
for $\theta_{12}^\ell=\pi/3$, $\theta_{13}^\ell=\pi/6$, $\theta_{23}^\ell=\pi/4$, and $\delta^\ell=0$. Comparing to the experimental bounds yields a $Z'$ mass around 3 TeV. 

%%%%%%%%%%%%%%%%%%%%%%%
%%%%%%%%%%%%%%%%%%%%%%%%%%%%%%%%  

\subsection{$\mu-e$ conversion in nuclei}
In this sector, we consider a hypothetical process, called $\mu\to e$ conversion in nuclei, in which
negative muons are captured in a target of atomic nuclei, such as Titanium (Ti), Aluminum (Al) or Gold (Au), to form muonic atoms. The muon then converses into an electron in the nuclear field without creating a neutrino. There are a number of experiments that have been built or planned to built to search for the process's signals, for instance TRIUMP \cite{Honecker:1996zf}, SINDRUM-II \cite{Bertl:2006up}, and 
COMET \cite{Kuno:2013mha}. The current experimental limits on the branching ratios are $4.3\times 10^{-12}$
at TRIUMF for Titanium target and $7\times 10^{-13}$ for Gold target by SINDRUM-II. Furthermore, the goal of the future experiment COMET is to probe $\mu-e$ conversion signals with sensitivity about $10^{-16}$.
	
In the considering model, the $\mu-e$ conversing ratio can be calculated from a Lagrangian, which has same 
form as (\ref{Semi1})
\bea
	\mathcal{L}_{\mu- e}^I= -\fr{4G_F}{\sqrt{2}}\left[
	h_{LL}^{q}\left(\bar{\mu}\ga^\mu P_L e \right)\left(\bar{q}\ga_\mu P_L
	q\right)+ h_{LR}^{q}\left(\bar{\mu}\ga^\mu P_L e \right)\left(\bar{q}\ga_\mu
	P_R q\right)\right]+H.c.,
	\label{MueConv}\eea
in which 
\be h_{LL}^{q}=\fr{g\sqrt{2}(2+c_{2W})\Ga_{\mu
			e}^{lZ^\prime}}{24G_Fm_{Z^\prime}^2c_W\sqrt{1+2c_{2W}}},\hs  
	h_{LR}^{q}=\fr{g\sqrt{2}s_W^2 \Ga^{l Z^\prime}_{\mu
			e}\eta_q}{12G_Fm_{Z^\prime}^2c_W\sqrt{1+2c_{2W}}}.\ee
The conversing ratio (CR), which is obtained after normalizing to the
	total nuclear capture rate $\omega_{capt}$, can be simply expressed as
	\bea
	\nonumber
	\mathrm{CR}(\mu^-A \rightarrow e^-A) &=&
	\fr{4G_F^2}{\omega_{capt}}\left|\left(2h_{LL}^u+2h_{LR}^u+h_{LL}^d+h_{LR}^d\right)V^{(p)}
	~~~~~~~~~~~~~~~~~~~~~~~~\right.\\
	&&\left.+\left(h_{LL}^u+h_{LR}^u+2h_{LL}^d+2h_{LR}^d\right)V^{(n)}\right|^2,
	\eea
where $V^{(n)}$ and $V^{(p)}$ are the overlap integrals and $\omega_{capt}$ is the total capture rate.
For the cases of $_{22}^{48}\text{Ti}$, $_{13}^{27}\text{Al}$, $_{79 }^{197}\text{Au}$, and
	$_{82 }^{208}\text{Pb}$ nuclei, they are given in Table \ref{NuclParam}
	\cite{mu-e-conversion1a}. 
%
%%%%%%%%%%%%%%%%%%%%%%%%%%%%%%%%%
	\begin{table}[b]
		\begin{center}
			\begin{tabular}{|cccc|}
				\hline \hline
				\rule[0.15in]{0cm}{0cm}{\tt $\mathcal{N}$} & $V^{(p)}\;m_{\mu}^{-5/2}$ &
				$V^{(n)}\;m_{\mu}^{-5/2}$ & $\omega_{\rm capt}~(10^{6}\,\text{s}^{-1})$ \\
				\hline
				\rule[0.25in]{0cm}{0cm}$^{48}_{22}$Ti  & 0.0396 & 0.0468 & 2.590\\
				\rule[0.25in]{0cm}{0cm}$^{27}_{13}$Al  & 0.0161 & 0.0173 & 0.7054\\
				\rule[0.25in]{0cm}{0cm}$^{197}_{79}$Au & 0.0974 & 0.146  & 13.07 \\
				\rule[0.25in]{0cm}{0cm}$^{208}_{82}$Pb & 0.0834 & 0.128  & 13.45 %\\
				%\rule[0.005in]{0cm}{0cm}&   & &
				\\\hline\hline
			\end{tabular}
			\caption{Nuclear parameters related to $\mu-e$ conversion in
				$_{22}^{48}\text{Ti}$, $_{13}^{27}\text{Al}$, $_{79 }^{197}\text{Au}$ and
				$_{82 }^{208}\text{Pb}$.}
			\label{NuclParam}
		\end{center}
	\end{table}%
	
\begin{figure}[t]
\begin{center}
\begin{tabular}{cc}
\includegraphics[width=7.5cm,height=6.5cm]{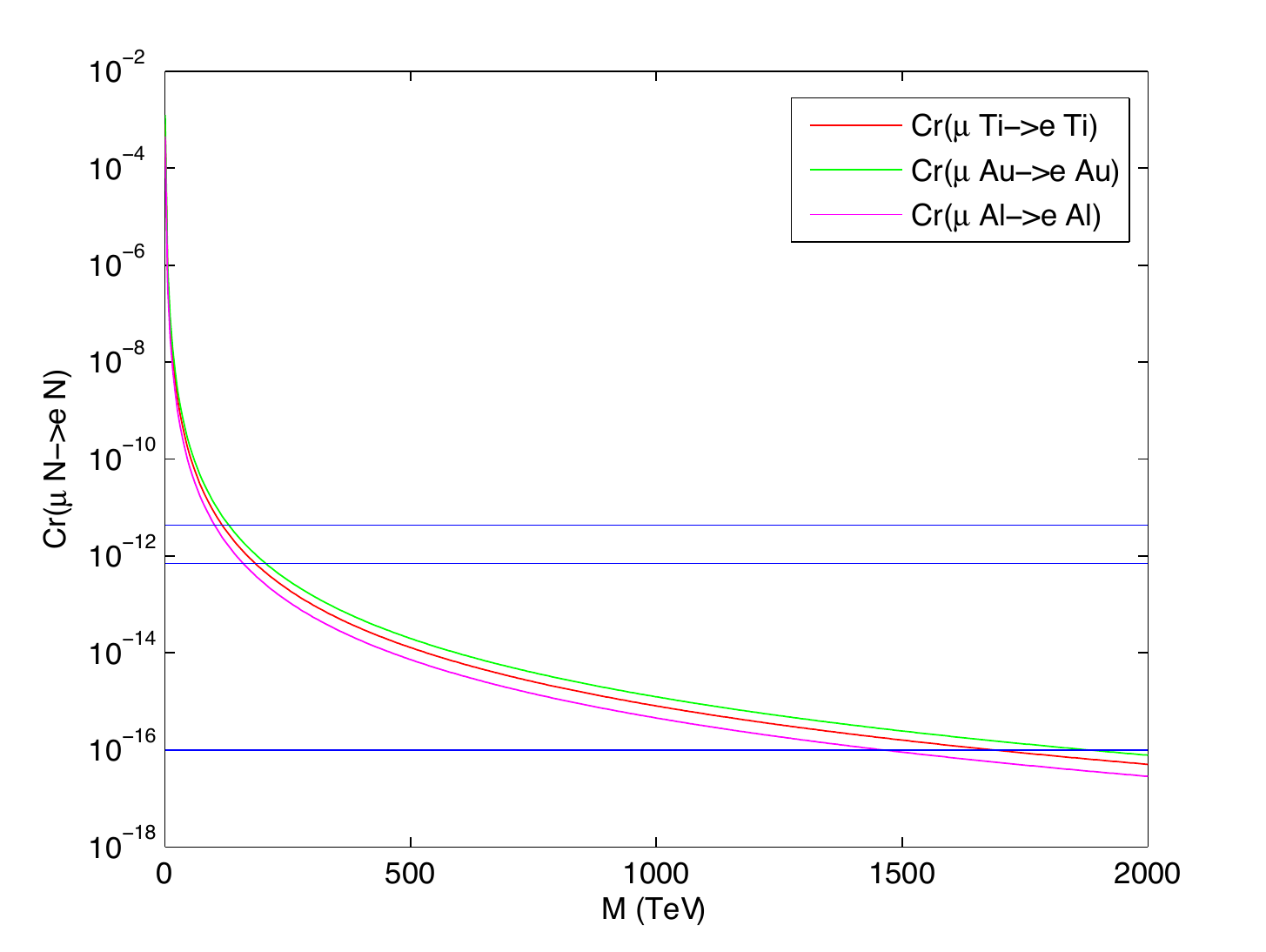} &
\includegraphics[width=7.5cm,height=6.5cm]{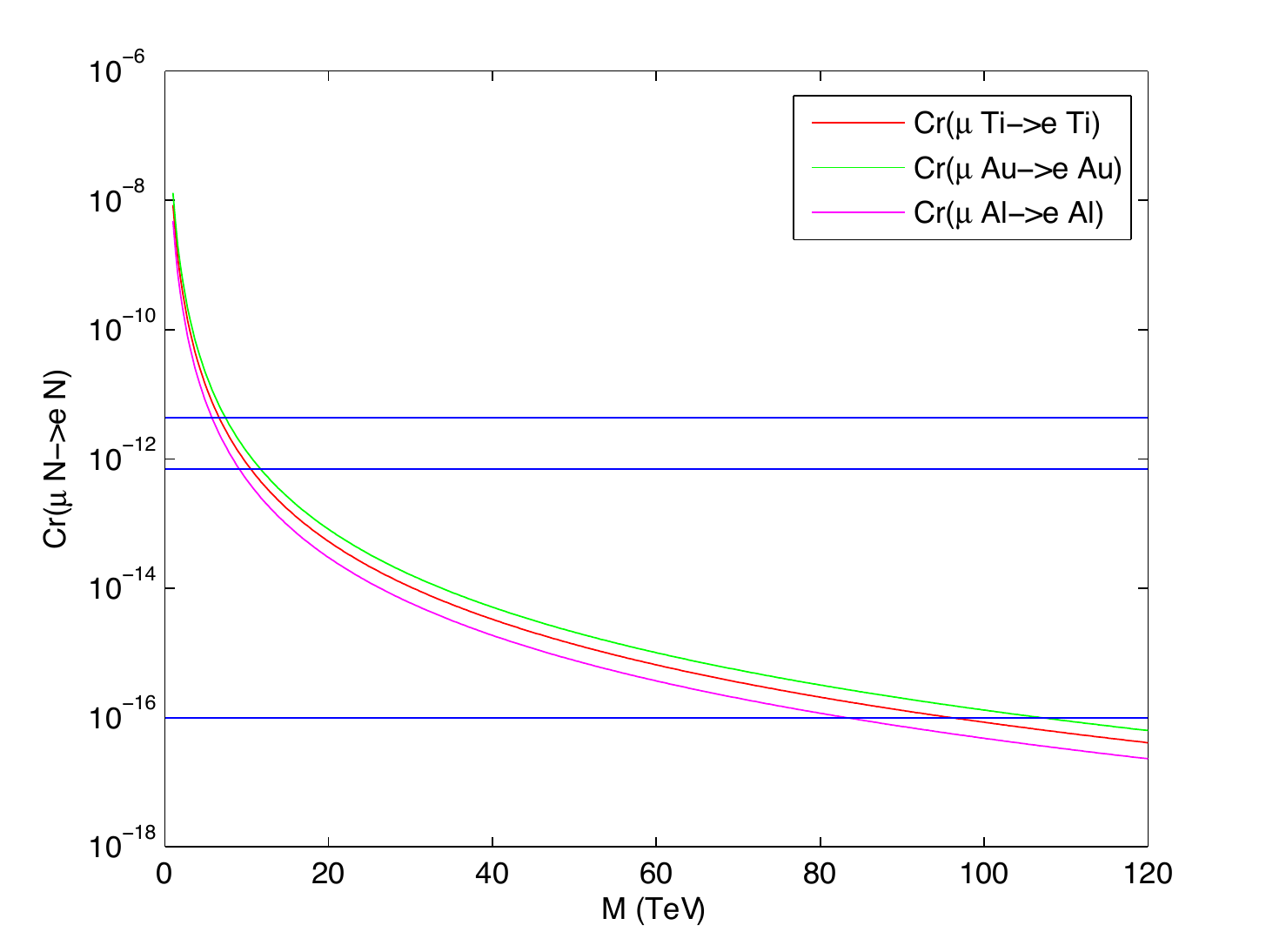}
\end{tabular}
 \caption{The $\mu\to e$ conversing ratio $\mathrm{Br}(\mu~N\to e~N)$ versus the new neutral 
  gauge boson mass $m_{Z'}\equiv M$, for different nuclei: \textit{i)} $^{48}_{22}$Ti (red line),
  \textit{ ii)} $^{27}_{12}$Al (magenta line), and
  \textit{iii)} $^{197}_{79}$Au (green line). The three blue lines corresponding to 
  $\mathrm{Br}(\mu ~Ti\to e~Ti)\leq 4.3\times 10^{-12}$ \cite{Honecker:1996zf}, 
  $\mathrm{Br}(\mu ~Au\to e~Au)\leq 7.0\times 10^{-13}$,
and $\mathrm{Br}(\mu ~Al\to e~Al)\leq 1.0\times 10^{-16}$ are the SINDRUM-II current upper bounds \cite{Bertl:2006up} and the COMET expected sentisitvity \cite{Kuno:2013mha}, respectively. The left panel is created with $\theta_{12}^\ell=\pi/3$, $\theta_{13}^\ell=\pi/6$, 
$\theta_{23}^\ell=\pi/4$, whereas the right panel is produced for $\sin\theta_{12}^\ell=0.9936$, $\sin\theta_{13}^\ell=0.9953$, $\sin\theta_{23}^\ell=0.2324$, $\delta^\ell=1.10\pi$.}
\label{Mu-e}
\end{center}
\end{figure}

We depict in Fig. \ref{Mu-e} the $\mu-e$ conversion ratios in the nuclei of Titanium, Aluminum, and Gold as functions of the new gauge boson mass $m_{Z'}$ for the same sets of parameter values, $\theta_{12}^\ell=\pi/3$, 
  $\theta_{13}^\ell=\pi/6$, $\theta_{23}^\ell=\pi/4$, $\delta^\ell=0$ (left panel) and 
  $\sin\theta_{12}^\ell=0.9936$, $\sin\theta_{13}^\ell=0.9953$, $\sin\theta_{23}^\ell=0.2324$, 
  $\delta^\ell=1.10\pi$ (right panel), used before concerning Fig. \ref{FigMu3e}. The present upper 
limits give stronger constraints on the new neutral gauge boson mass comparing with the other lepton-flavor violating processes considered before. For the first value set of $(\theta^\ell,\delta^\ell)$, consistency with the experimental results yields $m_{Z'}\geq 116.7$ TeV carried with Titanium target and $m_{Z'}\geq 204.5$ TeV with Gold target, respectively. Moreover, the planned experiment COMMET with sensitivity $10^{-16}$ is possible to probe the conversion signals as long as the new gauge boson is not heavier than about $1468.9$ TeV. If the lepton mixing parameters are taken as the second set, three lower/upper bounds obtained above are replaced by $m_{Z'}\geq 6.7$, 11.7, and 84.1 TeV. This second cause quite agrees with the collider searches. 
 %%%%%%%%%%%%%%%%%%%%%%%%%%%%%%%
 
Finally, we introduce here the viable regions of the branching ratios for the cases of Titanium, Aluminum, and Gold when varying the lepton mixing parameters in the allowed ranges, $\theta_{ij}^\ell$ in $[0,\pi/2]$ and $\delta$ in $[0,2\pi]$:
	%%%%%%%%%%%%%%%%%%%%%%%%%%%%%%%%%%%%%%%%%%%%
	\begin{eqnarray}
	&&0\leq {\rm Br}(\mu~{\rm Ti}\to e~{\rm Ti})\leq 1.9\times 10^{-3}\left(\frac{1\ {\rm TeV}}{m_{Z'}}\right)^4,\\
	&&0\leq {\rm Br}(\mu~{\rm Au}\to e~{\rm Au})\leq 3.0\times 10^{-3}\left(\frac{1\ {\rm TeV}}{m_{Z'}}\right)^4,\\
	&&0\leq {\rm Br}(\mu~{\rm Al}\to e~{\rm Al})\leq 1.1\times 10^{-3}\left(\frac{1\ {\rm TeV}}{m_{Z'}}\right)^4.
\end{eqnarray}
Here the maximal value occurs at 
$(\sin\theta_{12}^\ell,\sin\theta_{13}^\ell,\sin\theta_{23}^\ell,\delta^\ell)=(0.7072,0.000,0.8614,0.03\pi)$, 
while the minimal value occurs at $(\sin\theta_{12}^\ell,\sin\theta_{13}^\ell,\sin\theta_{23}^\ell,\delta^\ell)=(0.9758,1.0000,0.8658,1.62\pi)$, respectively.

\subsection{Constraining nonstandard neutrino interactions}

Let us study the phenomenological consequences of nonstandard neutrino interactions (NSIs) given in (\ref{dh3}), (\ref{dh4}), (\ref{dh5}), and (\ref{dh6}).
For convenience we write down the effective operators responsible for the NSIs as
	\cite{NSI1,NSI2, NSI3, NSI4}
	\bea
	\mathcal{L}_{\rm{NSI}}=-2\sqrt{2}G_F \epsilon^{fC}_{\al \beta}\left(\bar{\nu}_\al
	\ga_\mu P_L \nu_\beta \right) \left(\bar{f}\ga^\mu P_C f \right),
	\eea
where $\al, \beta$ denote the neutrino flavors $e$, $\mu$, and $\tau$. $P_C$ stands for the chiral projectors $P_L,P_R$. And, $f$ is the standard model first generation fermion $(e,u,d)$. The NSIs can affect neutrino oscillations in multiple ways, in the production, dectection and propagation of neutrinos. 

In this work, we consider the neutrino propagation in matter with the NSIs, assuming no effect of production and detection with the NSIs. The NSIs effect the neutrino propagation via coherent forward scattering in Earth matter. The Hamiltonian, which governs the propagation of neutrino flavor states in matter including the NSIs, is written as follows
	\bea
	\hat{H}=\fr{1}{2E}\left[U Diag(m_1^2,m_2^2,m_3^2)U^\dag +Diag(A,0,0)+A\epsilon^m
	\right],
	\label{Haeff}\eea
where $E$ is neutrino energy, $U$ is lepton mixing matrix, and $m_i$ are neutrino masses. Furthermore, $A=2\sqrt{2}EG_FN_e$ is the effective matter potential that is driven by ordinary charged-current weak interaction with electron. $N_e$ is the electron density along the neutrino trajectory. The matrix $\epsilon^m$ has a form as
	\bea 
	\epsilon^m=\left(  \begin{array}{ccc} 
		\epsilon_{ee}& \epsilon_{e\mu}&\epsilon_{e \tau}\\
		\epsilon_{e\mu}^* & \epsilon_{\mu \mu}& \epsilon_{\mu \tau} \\
		\epsilon_{e \tau}^* & \epsilon_{\mu \tau}^* & \epsilon_{\tau \tau} \\
	\end{array}\right),
	\eea
where $\epsilon_{\al \beta}$ is called a matter NSI parameter and is defined as
	\bea
	\epsilon_{\al \beta}=\sum_{f,C}\epsilon_{\al \beta}^{fC}\fr{N_f}{N_e},
	\eea 
where $N_f$ is the number density of a fermion of type $f$. In the considering model, the form of $\epsilon_{\al \beta}^{fC}$ can be obtained as follows
	\bea
	&& \epsilon_{\al \beta}^{eL}=\fr{\Ga_{\al \beta}^{\nu Z^\prime}\Ga_{ee}^{e
			Z^\prime}}{2\sqrt{2}G_F m_{Z^\prime}^2},\hs
			\epsilon_{\al
		\beta}^{eR}=\fr{\Ga_{\al \beta}^{\nu Z^\prime}}{2\sqrt{2}G_F
		m_{Z^\prime}^2}\left[\fr{g s_W^2}{c_W \sqrt{1+2c_{2W}}} \right],\nonumber \\
	&& \epsilon_{\al \beta}^{qL}=-\fr{\Ga_{\al \beta}^{\nu Z^\prime}}{2\sqrt{2}G_F
		m_{Z^\prime}^2}\left[\fr{g(2+c_{2W}) }{6c_W \sqrt{1+2c_{2W}}} \right], \label{EpsilonNon} \\
	&&\epsilon_{\al \beta}^{qR}=-\fr{\Ga_{\al \beta}^{\nu Z^\prime}}{2\sqrt{2}G_F
		m_{Z^\prime}^2}\left[\fr{gs_W^2 }{3c_W \sqrt{1+2c_{2W}}} \right].\nn
	\eea 
	
The effective Hamiltonian in (\ref{Haeff}) governs the neutrino propagation in matter due to the NSIs, hence varying the neutrino oscillation probabilities in comparison to the normal case \cite{Proba1}. However, at present, there is no evidence for the NSI associated with the experimental data of neutrino oscillations. The latest constraints on the NSIs from the global analysis of oscillation data can be found in \cite{Esteban:2018ppq}, which give the most stringent constraints on $\epsilon^{u}_{\alpha\beta}$ and $\epsilon^{d}_{\alpha\beta}$, namely $\epsilon^{u}_{\alpha\beta}$ bounded 
in the range $[-0.013,0.014]$ or  $[-0.012,0.009]$ while $\epsilon^{d}_{\alpha\beta}$ bounded in the range $[-0.012,0.009]$ and  $[-0.011,0.009]$, according to the cases of coherent data excluded or included, respectively.
	
From (\ref{EpsilonNon}), we roughly estimate 
	\begin{equation}
	\left|\epsilon_{\al \beta}^{fC}\right|\sim \frac{1}{2\sqrt{2}G_Fm_{Z'}^2}\simeq 3.0\times 10^{-2}
	\left[\frac{1\ {\rm TeV}}{m_{Z'}}\right]^2,
	\end{equation}
which is at order $10^{-2}$, $10^{-4}$, and $10^{-6}$ for $m_{Z'}=1$, 10, and 100 TeV, respectively. The predicted values lie within the experimental limits but do not give significant constraints on the model parameters. Comparing to the $\epsilon^{u}_{\alpha\beta}$ and $\epsilon^{d}_{\alpha\beta}$ bounds, it is hard to probe the nonstandard neutrino interactions in the model for $m_{Z'}>3$ TeV.
	
\subsection{LHC dilepton and dijet searches}

Since the new neutral gauge boson $Z'$ directly interacts with ordinary quarks ($q$) and leptons ($l$), the new physics processes $pp\to Z'\to f\bar{f}$ for $f=q,l$ exist at the LHC dominantly contributed by $Z'$. 

The cross-section for producing a dilepton or diquark final state can be computed with the aid of the narrow width approximation \cite{Accomando:2010fz}, 
\be \sigma(pp\to Z'\to f \bar{f})=\fr 1 3 \sum_q \fr{dL_{q\bar{q}}}{dm^2_{Z'}}\hat{\sigma}(q\bar{q}\to Z')\mathrm{Br}(Z'\to f\bar{f}),  \ee  
where the parton luminosities $dL_{q\bar{q}}/dm^2_{Z'}$ at the LHC $\sqrt{s}=13$ TeV can be found in \cite{Martin:2009iq}. The partonic cross-section and branching ratio $\mathrm{Br}(Z'\to f\bar{f})=\Ga(Z'\to f\bar{f})/\Ga_{Z'}$ are 
\bea
&&\hat{\sigma}(q\bar{q}\to Z')=\fr{\pi g^2}{12c^2_W}[(g^{Z'}_V(q))^2+(g^{Z'}_A(q))^2],\\
&& \Ga(Z'\to f\bar{f})=\fr{g^2m_{Z'}}{48\pi c^2_W}N_C[(g^{Z'}_V(f))^2+(g^{Z'}_A(f))^2],\\
&& \Ga_{Z'}=\fr{g^2m_{Z'}}{48\pi c^2_W}\sum_{f} N_C [(g^{Z'}_V(f))^2+(g^{Z'}_A(f))^2],\eea where $N_C$ is the color number of $f$ and assuming that ${Z'}$ decays only to fermions. Indeed, it is easily verified that the other $Z'$ decay channels such as to ordinary Higgs and gauge bosons give small contribution to the total $Z'$ width.

\begin{figure}[t]
\bc
\includegraphics[scale=0.7]{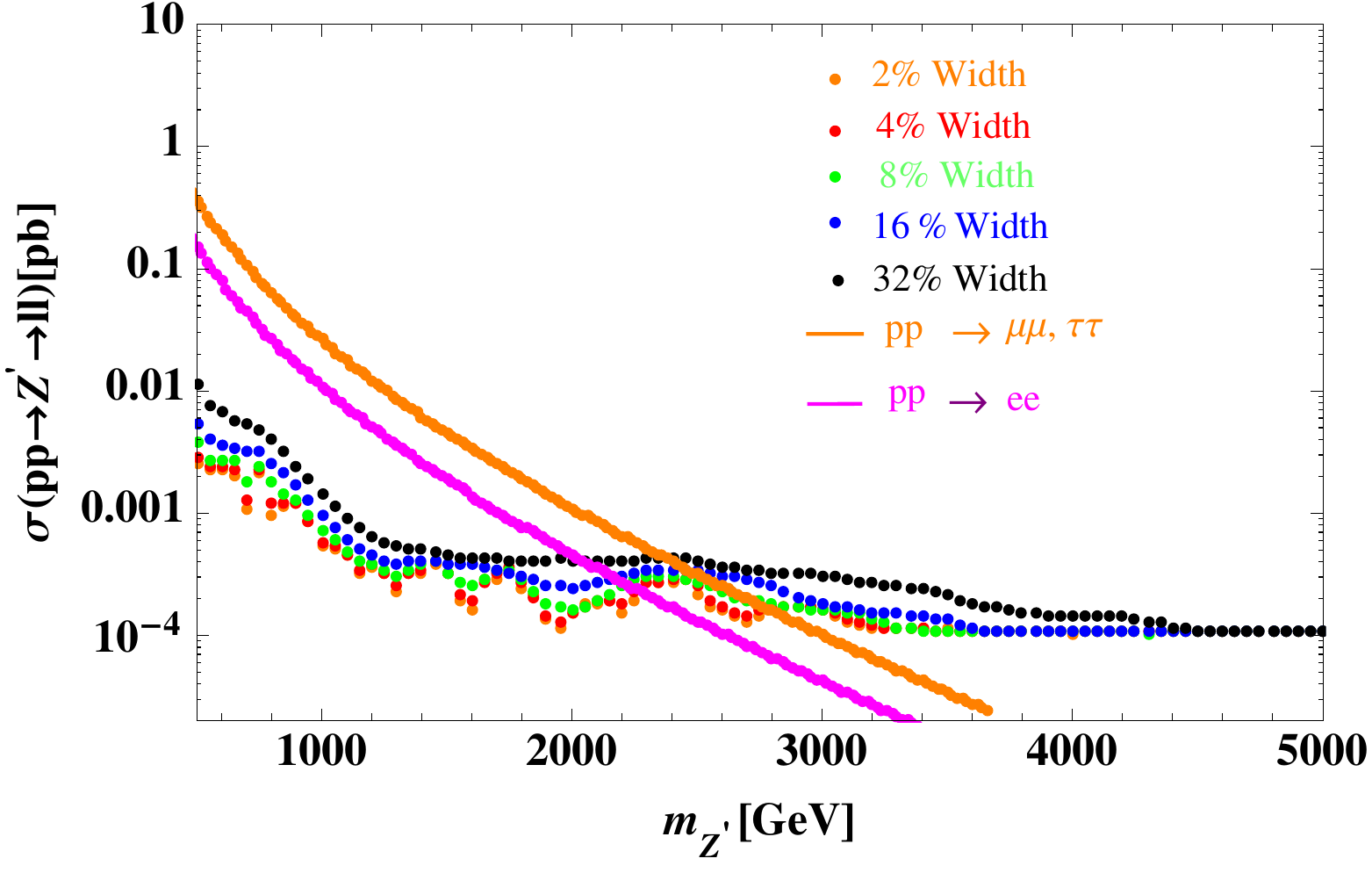}
\caption{\label{diadd} Dilepton production cross-section as a function of the new neutral gauge boson mass. The dotted lines are observed limits for different widths extracted at the resonance mass of dilepton, using 36.1 fb$^{-1}$ of $pp$ collision at $\sqrt{s}=13$ TeV by the ATLAS experiment \cite{Aaboud:2017buh}.}
\ec
\end{figure}

In Fig. \ref{diadd}, we show the cross-section for dilepton final states $l=e,\mu,\tau$. The experimental searches by the ATLAS \cite{Aaboud:2017buh} yield negative signals for new events of high mass, which transform to the lower limit for $Z'$ mass, $m_{Z'}>2.25$ and $2.8$ TeV, according to $ee$ and $\mu\mu(\tau\tau)$ channels, respectively. The last bound agrees with the highest invariant mass of dilepton hinted by the ATLAS. It is noteworthy that the $ee$ and $\mu\mu(\tau\tau)$ signal strengths are separated, which can be used to approve or rule out the flipped 3-3-1 model. 

Furthermore, the dijet production cross-section for $q=u,d$ can be evaluated by comparing the $Z'l\bar{l}$ and $Z'q\bar{q}$ couplings. Indeed, since $\Gamma(Z'\to u\bar{u}) =\Gamma(Z'\to d\bar{d}) \simeq 2.3 \Gamma(Z'\to e\bar{e}) \simeq \Gamma(Z'\to \mu \bar{\mu})=\Gamma(Z'\to \tau \bar{\tau})$, this leads to $\sigma(pp\to Z'\to q\bar{q})\sim \sigma(pp\to Z'\to l\bar{l})$. Because the current bound on dijet signals is less sensitive than the dilepton \cite{Aaboud:2017yvp},  the corresponding $Z'$ mass limit is quite smaller than that obtained from the dilepton, which is not included. In sort, in the present model, the dijet sinals predicted are negligible, given that the dilepton bound applies for $Z'$.               
	
\subsection{Dark matter}

The model contains two kinds of dark matter candidates: (i) the fermion triplet $\xi$ which is unified with the standard model lepton doublet $(\nu_{1L}\ e_{1L})$ in the $SU(3)_L$ sextet and (ii) the scalar that is either $\rho_3$ or a combination (called $D$) of $\chi_2$ and $S_{23}$, whereas the remaining combination of $\chi_2$ and $S_{23}$ is the Goldstone boson of the $Y$ gauge boson. The candidate $D$ transforms as a standard model doublet, which interacts with $Z$. This gives rise to a large direct dark matter detection cross-section that is already ruled out \cite{Barbieri:2006dq}. The singlet candidate $\rho_3$ can fit the relic density and detection experiments, which has been studied extensively \cite{Dong:2014wsa,Huong:2018ytz}. The fermion candidate $\xi$ is a new observation of this work\footnote{In the minimal dark matter, such candidate was ruled out because a stability mechanism for dark matter was not appropriately taken \cite{Cirelli:2005uq}.}.

Note that at the tree-level, the components of $\xi$ triplet have degenerate masses, already obtained as $m_\xi$. However, the loop effects of gauge bosons can make the $\xi^\pm$ mass bigger than the $\xi^0$ mass by $m_{\xi^\pm}-m_{\xi^0}=166$ MeV \cite{Cirelli:2005uq}. Therefore, $\xi^0$ can be regarded as a LWP responsible for dark matter.  

In the early universe, the dark matter candidate $\xi^0$ can (co)annihilate into the standard model particles that set its abundance. Generalizing the result in \cite{Cirelli:2005uq}, we obtain the annihilation cross-section,
\bea \langle \sigma v\rangle &\simeq& \fr{37g^4}{96\pi m^2_\xi}\crn
&\simeq& \left(\fr{\al}{150\ \mathrm{GeV}}\right)^2\left(\fr{2.86\ \mathrm{TeV}}{m_\xi}\right)^2, \eea where $(\al/150\ \mathrm{GeV})^2\simeq 1$ pb. Comparing to the observation, we have $\Omega_{\xi}h^2\simeq 0.1\ \mathrm{pb}/\langle \sigma v\rangle \simeq 0.11$ \cite{Tanabashi:2018oca}, which implies $m_\xi\simeq 2.86$ TeV.

The dark matter $\xi^0$ can scatter off nuclei causing observed effects in the direct detection experiments. At the tree-level, it does not interact with quarks confined in nucleons, since $T_3(\xi^0)=Y(\xi^0)=0$. The direct detection cross-section starts from the one-loop level, contributed by $W,h$ and $\xi^\pm$, leading to $\sigma_{\mathrm{SI}}\simeq 1.2\times 10^{-45}\ \mathrm{cm}^2$ \cite{Cirelli:2009uv}, in agreement with the experiment for the heavy dark matter mass $m_\xi\simeq 2.86$ TeV \cite{Aprile:2017iyp}.

The above evaluation is valid when the new gauge $(Z')$ and Higgs $S_{33}$ portals as well as the 3-3-1 model new particles are heavier than the dark matter, so that the dark matter observables are governed by the standard model particles. Alternatively, since both $Z'$ and $S_{33}$ couple to $\xi^0$, this dark matter can annihilate to the standard model Higgs, weak bosons and top quark as well as the appropriate new particles of the 3-3-1 model. In this case, the $Z'$ and $S_{33}$ resonances set the dark matter density and direct detection cross-section \cite{Dong:2014wsa,Dong:2015rka,Huong:2018ytz}. That said, we have two viable regimes responsible for the dark matter mass $m_{\xi^0}=\fr{1}{2} m_{Z'}$ and $m_{\xi^0}= \fr 1 2 m_{S_{33}}$, provided that $m_{Z'}$ and $m_{S_{33}}$ are separated. By contrast, these regimes are coincided. The $Z'$ mass bound tells us that the dark matter mass is at TeV or higher scale. This mass is easily evaded the direct detection \cite{Aprile:2017iyp}.               

\section{\label{conl} Conclusion}

The discovery of the flipped 3-3-1 model has changed the current research of the 3-3-1 model. Indeed, the flavor nonuniversality that is now associated with leptons due to the anomaly cancellation provides the need for realizing the type I and II seasaw mechanism and the matter parity naturally. We have shown that the neutrino masses are produced, yielding the seesaw scales $\kappa$ at eV and $\mathcal{M}$ at $10^{14}$ GeV. Whereas, the dark matter candidate may be a fermion or scalar that has the mass in the TeV scale.         
	
Because of the lepton generation nonuniversality, the charged lepton flavor violating processes and nonstandard neutrino interactions arise at the tree-level due to the exchange of $Z'$ boson. We have made a systematic search for such processes and found that the 3-3-1 breaking scale is coincided with those obtained from the LHC dilepton and dark matter constraints, where the $Z'$ portal governs the dark matter observables. This is an advantage over the ordinary 3-3-1 models since their FCNC constraints often do not coincide with the collision bounds \cite{Dong:2014wsa,Dong:2015dxw,Dong:2017ayu}.

All the results arise from the gauge symmetry principles. Hence, the flipped 3-3-1 model is very predictive that deserves further studies.         

\section*{Acknowledgments}
	
P.V.D. and D.T.H. would like to thank Prof. Osamu Yasuda for the discussions on nonstandard interactions of neutrinos. This research is funded by Vietnam National Foundation for Science and Technology Development (NAFOSTED) under grant number 103.01-2017.05.
	
\bibliography{combine}
	
\end{document}